\documentclass[sn-mathphys-num]{sn-jnl}
\usepackage[utf8]{inputenc}

\usepackage{graphicx}
\usepackage{amsmath,amssymb,amsfonts}%
\usepackage{hyperref}

\usepackage{booktabs}
\usepackage{flafter}

\usepackage{caption}
\usepackage{subcaption}

\usepackage{pgfplots}
\graphicspath{ {./figures/} }

\usepackage{amsthm}%
\usepackage{mathrsfs}%
\usepackage[title]{appendix}%
\usepackage{xcolor}%
\usepackage{textcomp}%
\usepackage{manyfoot}%
\usepackage{booktabs}%
\usepackage{algorithm}%
\usepackage{algorithmicx}%
\usepackage{algpseudocode}%
\usepackage{listings}%

\newcommand{\Msss}{m}

\newcommand{\Stfs}{k}

\newcommand{\Pls}{\omega}
\newcommand{\Npls}{\nu}
\newcommand{\Ang}{\alpha}
\newcommand{\Scl}{s}
\newcommand{\Hds}{w}
\newcommand{\Vds}{h}
\newcommand{\Num}{n}
\newcommand{\Rsd}{_{\text{res}}}
\newcommand{\Dnm}{_{\text{dyn}}}
\newcommand{\Spr}{_{\text{sp}}}
\newcommand{\Csc}{_{\text{cs}}}

\begin{document}
        
\title{Generative inverse design of multimodal resonant structures for locally resonant metamaterials}

\author[1]{\fnm{Sander} \sur{Dedoncker}}
\equalcont{These authors contributed equally to this work.}
\author[2]{\fnm{Christian} \sur{Donner}}
\equalcont{These authors contributed equally to this work.}
\author[2]{\fnm{Raphael} \sur{Bischof}}
\author[1]{\fnm{Linus} \sur{Taenzer}}
\author*[1]{\fnm{Bart} \sur{Van Damme}}\email{bart.vandamme@empa.ch}

\affil*[1]{\orgdiv{Laboratory for Acoustics/Noise Control}, \orgname{Empa, Materials Science and Technology}, \orgaddress{\street{Ueberlandstrasse 129}, \city{8600 D\"bendorf}, \country{Switzerland}}}

\affil[2]{\orgname{Swiss Data Science Center}, \orgaddress{\street{Andreasstrasse 5}, \city{8092 Z\"urich}, \country{Switzerland}}}
    
\abstract{In the development of locally resonant metamaterials, the physical resonator design is often omitted and replaced by an idealized mass-spring system. This paper presents a novel approach for designing multimodal resonant structures, which give rise to multi-band gap metamaterials with predefined band gaps. Our data science-based method uses a conditional variational autoencoder to identify non-trivial patterns between design variables of complex-shaped resonators and their modal effective parameters. After training, the cost of generating designs satisfying arbitrary criteria -- frequency and mass of multiple modes -- becomes negligible. An example of a resonator family with six geometric variables and two targeted modes is further elaborated. We find that the autoencoder performs well even when trained with a limited dataset, resulting from a few hundred numerical modal analyses. The method generates several designs that very closely approximate the desired modal characteristics. The accuracy of the best designs, proposed by the auto-encoder, is confirmed in tests of 3D-printed resonator prototypes. Further experiments demonstrate the close agreement between the measured and desired dispersion relation of a sample metamaterial beam.}
    
\keywords{metamaterial vibration resonator generative design}

\maketitle

\section{Introduction}

In recent years, metamaterials have garnered much interest in several engineering fields because of their extraordinary potential to break with traditional design goals \cite{Surjadi2019}. By virtue of a carefully designed mesostructure, metamaterials display exotic macroscopic properties that do not exist in known bulk materials. In noise and vibration engineering, in particular, locally resonant metamaterials (LRMs) have come to the fore as a promising technology \cite{Liu2000a,Ma2016}. By exploiting the phenomenon of resonance, LRMs can direct the energy present in acoustic waves and structural vibrations. This has many applications, ranging from energy harvesting to suppressing vibrations or noise \cite{Jimenez2021,Deymier2013a,dalela2022review}. 

Since their conception, LRMs have been realized in many different ways, going from basic material inclusions \cite{Liu2000a} or plates with cantilever resonators \cite{Claeys2013} to more advanced bio-inspired designs \cite{Murer2023,DalPoggetto2023}. Ad-hoc design approaches, often based on simplified metamaterial models, appear to be the most common. Where computational resources allow it, shape and/or topology optimization of LRM band gaps has been a popular research topic \cite{Matsuki2014,Krushynska2014,Yang2016,Roca2019,Jung2020,Huo2023,hedayatrasa2018optimization,giraldo2022design}.

Machine-learning (ML) inverse design has also gained attention in recent years, where the costly finite element (FE) analysis in the design pipeline is replaced by cheaper data-driven surrogate models. Generative models such as variational autoencoders (VAEs)~\cite{Kingma2013Auto-EncodingBayes} or generative adversarial networks~\cite{Goodfellow2014GenerativeNets} have recently been used for the design of molecules and metamaterials ~\cite{Gomez-Bombarelli2018AutomaticMolecules, Wang2020DeepSystems, Gurbuz2021GenerativeMetamaterials,vanmastrigt2022machine}. While these design strategies still require optimization guided by some learned surrogate model, on-demand design was also addressed ~\cite{Ma2018a, Pathak2020DeepGenerator,zheng2021data}, i.e. the desired properties of the design are already provided to the model. Several recent overview papers highlight the strength of these methods, showing that most of the work in the field of metamaterials for structural dynamics is done for one- and two-dimensional structures, leaving much to be explored on the topic of three-dimensional designs~\cite{jin2022intelligent,he2023AIreview,cerniauskas2024machine}.

The main goal of our present work is to further facilitate LRM design by applying generative ML methods. More concretely, a practical approach is presented to find suitable resonators that lead to multiple predefined band gaps for bending waves in an elastic beam. To achieve this goal, we focus on the three-dimensional design of mechanical resonators. This is motivated by the fact that, in an LRM, the metamaterial behavior is determined in the first place by the properties of the individual resonators, rather than their particular spatial arrangement \cite{Deymier2013a, Liu2012a}. Similar resonators are assumed to produce similar band gaps even in dissimilar host structures. Furthermore, by combining resonator types or exploiting additional modes in a single resonator, multiple band gaps can be created \cite{Claeys2017, Murer2023, Giannini2023}. The efficiency of the band gaps, quantified by the imaginary component of the wave number and its frequency span, is defined by the reaction force the resonator exerts on the host structure. This can equivalently be expressed by the dynamic mass at the resonance frequency of interest. 

Finding a suitable geometry for a mechanical resonator with multiple predefined eigenfrequencies and modal masses is not straightforward. Geometrical and material properties affect the modal properties of a complex-shaped structure in a non-trivial way. Several recent publications~\cite{li2020designing,jin2022deep} show how deep learning can be used to define the non-linear relation between a metamaterial's geometry and band gap. Both examples rely on material removal from a flat plate and can therefore be considered to be 2D-optimizations. Several examples exist where ML is used to optimize three-dimensional structures to achieve predefined static mechanical properties~\cite{Kumar2020Inverse-designedMetamaterials,zheng2023unifying}. Generative design of three-dimensional dynamic structures using standard topology optimization requires many iterations and involves the risk of reaching a local minimum of the cost function, which only optimizes part of the multiple desired dynamic variables. As an alternative, we propose a data-driven approach handling the resonator design by an inexpensive surrogate model, as has e.g. been done for the design of mechanical metamaterials with a desired stiffness tensor~\cite{wang2020deep,bastek2023inverse,zheng2023unifying} and acoustic metamaterials~\cite{zhang2021accelerated}. We ultimately aim to create LRMs with multiple on-demand band gaps, induced by a single resonant structure. This design problem requires the optimization of the resonator's eigenfrequencies, fixing the location of the band gaps, but also its modal masses, since they determine the efficiency of the wave attenuation and the width of the band gap. 
\citet{wang2020deep} pointed out the strength of using properties of the latent space of a variational autoencoder in the design process of metamaterials. Extending this idea, our proposed method using a conditional variational autoencoder is able to provide hundreds of (potentially very different) designs in a very short time and with low computational effort, whereas classical topology optimization only delivers one design for each run. The generated structures can be ranked using the surrogate model, so that the designs best matching all requirements can be selected. 

The remainder of this paper discusses our approach as follows. In the first section, we lay out the elementary parts -- a suitable representation of resonator dynamics as well as a VAE network -- and show how they fit together in our method. In the next section, we consider a parametric family of resonators that complies with practical manufacturing constraints. Many designs are randomly sampled and converted to high-fidelity FE models, from which modal parameters are extracted. The generated set of input and output data is used to train the VAE model, which once trained can provide designs with predefined resonant properties. We then assess the network performance and accuracy and use it to find designs satisfying a predefined dynamic response. A few select designs are manufactured and studied in detail. We compare numerical predictions and experimental measurements of their modal parameters and the resulting metamaterial properties. 

\section{Inverse design problem for multi-band gap elastodynamic metamaterials}

\subsection{Modal properties of mechanical resonators with multiple degrees of freedom}

In order to create an effective LRM, the motion between the resonators and the host should be adequately coupled and the resonator's dynamic mass should be large \cite{Deymier2013a}. For the latter requirement, the resonating mass should be high and the tuning frequency needs to match the excitation frequency, so the dynamic mass is strongly amplified. The classic representation is a mass-spring resonator with a weightless attachment point. The dynamic mass in this case may be defined as the ratio of the force and the acceleration of the attachment point. For a system with mass $\Msss$, stiffness $\Stfs$ and natural pulsation $\Npls = \sqrt{\Stfs/\Msss}$, this becomes 
\begin{equation}
    \Msss\Dnm(\Pls) = m\frac{\Npls^2}{\Npls^2-\Pls^2},
\end{equation}
highlighting the importance of both the (physical) mass and the tuning frequency. In the presence of damping, the stiffness and therefore the natural pulsation can be represented by complex values, thereby avoiding an infinite dynamic mass.

In the present study, we consider the resonator as a three-dimensional continuous structure, which is an important step towards practical implementations. Its motion has infinitely many degrees of freedom and an associated direction defined by the reaction force exerted on the boundary condition. Despite this apparent complexity, under relatively mild assumptions, it is still possible to derive an equivalent model from the structure's modal parameters. An in-depth discussion can be found in the work of Girard and Roy \cite{Girard208}. Concentrating on beam-, and shell-like host structures -- where flexural waves dominate -- we only consider coupling through transverse translations and forces. Moreover, the (relatively small) contact surface between the resonator and the host is assumed not to deform. Finally, assuming a hysteretic material damping model for the resonator, its modes can be decoupled. Under these conditions, a single resonator can be represented by a parallel arrangement of mass-spring systems, each corresponding to one vibration mode: 
\begin{equation} \label{eq:dynmss}
    \Msss\Dnm(\Pls) = \sum_{i}\Msss_i\frac{\Npls_i^2}{\Npls_i^2-\Pls^2} + \Msss\Rsd.
\end{equation}
Here, the modal effective masses $\Msss_i$ replace a single lumped mass, the (complex) modal frequencies $\Npls_i$ replace the natural pulsation, and the residual mass $\Msss\Rsd$ is used to account for the mass that is not linked to any modes. In particular, we define $\Msss\Rsd$ such that the static mass $\Msss\Dnm(0)$ equals the physical mass, regardless of how many modes are truncated from the sum. If such a resonator is used for the creation of SLMs, each modal mass-pulsation couple contributes to a band gap. 

\subsection{Dispersion of locally resonant metamaterials}
In order to assess resonator performance in the context of LRMs, the dispersion diagram can be studied. Band gaps are expected around each resonator mode, where the width and efficiency relate to the modal mass. For the sake of simplicity, we present a beam-type metamaterial with multiple band gaps for flexural waves. To calculate the dispersion curves we use the method of Liu and Hussein, which is a transfer matrix approach analyzing the host-resonator unit cell \cite{Liu2012a}. In the unit cell analysis, the parallel mass-spring model of eq.~\eqref{eq:dynmss} can be used without further changes, assuming that there are multiple spring and mass combinations connected to the same point on the host beam.

The real part of the dispersion curves yields the wave number and thus the bending wave speed as a function of the excitation frequency, showing typical asymptotic anomalies in band gap regions. The LRM’s efficiency can be deduced from the imaginary part of the dispersion relation. The higher the imaginary wave number, the faster a wave decays.

\subsection{Generative inverse design for metamaterial resonators using machine learning}
Topology optimization of structures to achieve a predefined static or dynamic behavior is an active topic of research in mechanical engineering, and typically relies on iterative explicit finite element analyses of the structures under consideration.
To mitigate the computational cost of many iterations which are required to find a suitable resonator design for given tuning frequencies and modal masses, we train a data-driven surrogate model. It is cheap to evaluate, but still sufficiently accurate to map the resonator geometry to its dynamic properties. Here, we resort to a conditional variational autoencoder (cVAE). A variational autoencoder (VAE)~\cite{Kingma2013Auto-EncodingBayes} is a cyclic architecture, which has two parts: an encoder, which maps parameterized designs to a probabilistic latent space, and a decoder, which provides the approximate inverse map from the latent space back to the design parameters. Typically, the encoder and decoder are modeled by neural networks. Once the VAE is trained on some initial dataset, the user can generate new designs by sampling the latent space and then projecting the samples back to the design space via the decoder. In the case of the conditional VAE~\cite{Sohn2015LearningModels}, the user additionally provides desired properties for the design as input to the decoder. If the cVAE correctly learned relations between the requested parameters, the generated designs have properties that are close to the requested ones, allowing for efficient design without any post hoc optimization. Here we use a recently developed cVAE framework~\cite{Salamanca2023AugmentedStudy2} to demonstrate its effectiveness for targeted LRM design.

While cVAEs are a powerful tool for a wide range of inverse design problems, certain limitations have to be taken into account. Most importantly, the neural network implementation assumes (implicitly) that the modeled quantities exist in vector spaces and are related by continuous functions. Seen from the perspective of resonator design, it is not always trivial to identify such quantities. For instance, mode shapes or reaction force vectors have an arbitrary scale and orientation and are therefore not directly suited. Fortunately, as we saw before, we can reduce the relevant dynamics to two positive scalar parameters (mass and frequency) per mode.

Still, there remains one issue. Because modal analysis is essentially an eigenvalue decomposition, it can be assumed that the outcome varies continuously with e.g. geometric dimensions. However, different modes exist on different `branches' of the decomposition. It can occur that, for one design, branch A (e.g. a flexural mode) has a higher frequency than branch B (e.g. a torsional mode), whereas, for another design, it is the other way around. Because modes are typically ordered by frequency in the analysis, assigning modes to the continuous branches is not straightforward.  These discontinuous changes within the design space render classical (gradient-based) topological optimization for a combination of desired modal properties unpractical.


While addressing this problem fully is out of this work's scope, we try to mitigate its effects by the following two-step selection procedure. We first pick the first $N$ modes with the largest modal effective mass along the desired direction. Then we sort the selected modes by ascending frequency, e.g. the first mode is the mode with the lowest frequency among the $N$ modes with the largest mass.

In summary, our proposed method will use a cVAE to create an invertible mapping between, on the one hand, design parameters such as geometry, and on the other, the frequencies and modal effective masses of the `heaviest' modes. They are assumed to induce the most effective band gaps. The design parameters could also include material properties such as mass density or Young's modulus when this is technically feasible. For demonstration purposes, we limit ourselves to $N=2$, but in principle there is no limitation on the number of modes the VAE is trained for. 


\section{Implementation of a machine learning approach for the design of a multi-band gap LRM}

\subsection{Parametric resonator design} \label{ssec:rd}

For the resonator design, we target frequencies on the order of 10 to 1000\,Hz. A single geometry should show multiple resonance frequencies. The modal frequencies of simple geometries such as cantilever beams are highly correlated. We therefore investigate a family of shapes which can be described by a limited number of geometric parameters, but with sufficiently rich dynamics for the set goals. We furthermore prescribe that the resonators can be manufactured using a selective laser sintering machine. The maximum and minimum feature sizes should therefore be in the order of 100\,mm and 1\,mm. We assume that the fusion process results in a homogeneous and isotropic material with density 975\,kgm$^{-3}$, Young's modulus 2.0\,Gpa, and Poisson's ratio 0.3. These specific values were obtained by testing printed beam samples, and are consistent with the listed properties of the printing polymer (Nylon PA 12).

\begin{figure}[h]
    \centering
    \def\svgwidth{0.8\textwidth}	
\begingroup%
  \makeatletter%
  \providecommand\color[2][]{%
    \errmessage{(Inkscape) Color is used for the text in Inkscape, but the package 'color.sty' is not loaded}%
    \renewcommand\color[2][]{}%
  }%
  \providecommand\transparent[1]{%
    \errmessage{(Inkscape) Transparency is used (non-zero) for the text in Inkscape, but the package 'transparent.sty' is not loaded}%
    \renewcommand\transparent[1]{}%
  }%
  \providecommand\rotatebox[2]{#2}%
  \newcommand*\fsize{\dimexpr\f@size pt\relax}%
  \newcommand*\lineheight[1]{\fontsize{\fsize}{#1\fsize}\selectfont}%
  \ifx\svgwidth\undefined%
    \setlength{\unitlength}{1258.81494717bp}%
    \ifx\svgscale\undefined%
      \relax%
    \else%
      \setlength{\unitlength}{\unitlength * \real{\svgscale}}%
    \fi%
  \else%
    \setlength{\unitlength}{\svgwidth}%
  \fi%
  \global\let\svgwidth\undefined%
  \global\let\svgscale\undefined%
  \makeatother%
  \begin{picture}(1,0.52940991)%
    \lineheight{1}%
    \setlength\tabcolsep{0pt}%
    \put(0,0){\includegraphics[width=\unitlength,page=1]{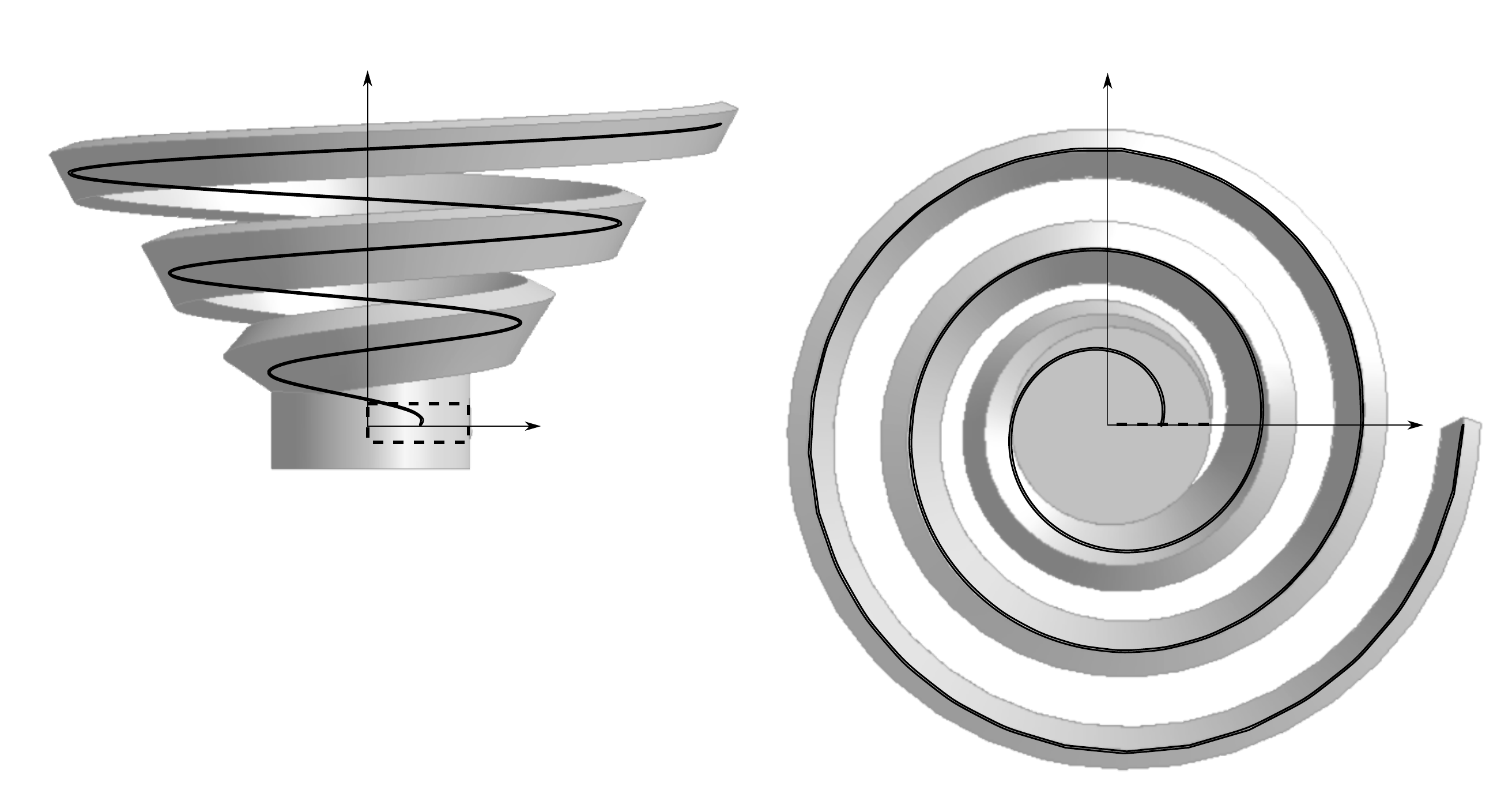}}%
    \put(0.25572726,0.46885226){\color[rgb]{0,0,0}\makebox(0,0)[lt]{\lineheight{1.25}\smash{\begin{tabular}[t]{l}$z$\end{tabular}}}}%
    \put(0.74514343,0.46649009){\color[rgb]{0,0,0}\makebox(0,0)[lt]{\lineheight{1.25}\smash{\begin{tabular}[t]{l}$y$\end{tabular}}}}%
    \put(0.35608391,0.2539898){\color[rgb]{0,0,0}\makebox(0,0)[lt]{\lineheight{1.25}\smash{\begin{tabular}[t]{l}$x$\end{tabular}}}}%
    \put(0.94072068,0.25516037){\color[rgb]{0,0,0}\makebox(0,0)[lt]{\lineheight{1.25}\smash{\begin{tabular}[t]{l}$x$\end{tabular}}}}%
    \put(0,0){\includegraphics[width=\unitlength,page=2]{spiral_resonator.pdf}}%
    \put(0.206563,0.13993372){\color[rgb]{0,0,0}\makebox(0,0)[lt]{\lineheight{1.25}\smash{\begin{tabular}[t]{l}10mm\end{tabular}}}}%
  \end{picture}%
\endgroup%

    \caption{Geometry of the resonator. The spiral arm is constructed by sweeping a rectangle (dashed line) along a curve (solid line). The arm is partially embedded in a solid cylinder that is aligned with the spiral axis.}
    \label{fig:res}
\end{figure}

To meet the specifications, we propose a resonator design that consists of 2 intersecting bodies (Figure \ref{fig:res}). The first body is a solid cylinder, which is intended to provide a flat, stiff, and invariant surface through which the resonator can be attached to the host. The cylinder has a radius of 10\,mm, is aligned along the $z$-axis and extends between $z=-5$\,mm and $z=10$\,mm. The second body is a parameterized spiral-shaped arm, which is intended to provide practical modal masses in the desired low frequency range. To construct the arm, we first define a parametric curve
\begin{equation}
  \left\{\begin{aligned}
  x&=\Hds\Spr\left(\frac{(\Scl\Spr-1)}{2\pi\Num\Spr}\Ang+1\right)\cos{\Ang} \\ 
  y&=\Hds\Spr\left(\frac{(\Scl\Spr-1)}{2\pi\Num\Spr}\Ang+1\right)\sin{\Ang} \\
  z&=\frac{\Vds\Spr}{2\pi\Num\Spr}\Ang
  \end{aligned}\right. \qquad 0\leq\Ang\leq2\pi\Num\Spr,
\end{equation}
where $\Hds\Spr=5$\,mm is the spiral `width', $\Vds\Spr$ is the spiral height, $\Scl\Spr$ is the spiral scaling factor, and $\Num\Spr$ is the number of turns. A rectangle, centered on the spiral, is swept along the curve. At the start, this rectangle has width $\Hds\Csc$ and height $\Vds\Csc$ and is contained by the $xz$-plane. As the rectangle is extruded, its orientation relative to the curve is fixed, but its dimensions change linearly with the angle $\Ang$. At the end of the curve, the rectangle becomes scaled by a factor $\Scl\Csc$.

We obtain a geometric family of resonators by independently varying the six free parameters ($\Vds\Spr$, $\Hds\Csc$, $\Vds\Csc$, $\Scl\Spr$, $\Num\Spr$ and $\Scl\Csc$). Considering the limitations of the manufacturing process, these parameters are bounded by the values in Table \ref{tab:bc}. In addition, to prevent self-intersecting or oversized geometry, the following two constraints are imposed:
\begin{equation}
  \begin{aligned}
  &\Hds\Csc\max(\Scl\Csc,1) \leq \Hds\Spr\frac{\Scl\Spr-1}{\Num\Spr} \\ 
  &\Hds\Csc\max(\Scl\Csc,1) + 2\Hds\Spr\Scl\Spr \leq 80\,\text{mm} 
  \end{aligned}
\end{equation}
		
\begin{table}[h]
\centering
    \caption{Resonator design parameters with their bound constraints.}
    \label{tab:bc}
    \begin{tabular}{lll}
        \toprule
        Parameter   				            &	Lower bound	&	Upper bound	\\
        \midrule
        Spiral height $\Vds\Spr$      	        &	0\,mm			&	30\,mm\\
        Spiral scaling factor $\Scl\Spr$        &  	2			&	8   \\
        Spiral number of turns $\Num\Spr$  	    &	1           &	4	\\
        Cross-section width $\Hds\Csc$ 	        &	2\,mm			&	10\,mm\\
        Cross-section heigth $\Vds\Csc$         &	2\,mm			&	4\,mm	\\
        Cross-section scaling factor $\Scl\Csc$ &	0.5			&	2	\\
        \bottomrule
    \end{tabular}

\end{table}

The advantages of graded spirals, a bio-inspired concept based on the cochlea, has attracted the attention of researchers in the field of metamaterials before~\cite{ma2014cochlear,zhao2019compact,poggetto2023cochlea}. They combine compactness with a wide range of accessible frequencies and a sufficiently large design space. The possibility to quickly and efficiently optimize the geometry to desired specifications opens the way to a variety of applications.

\subsection{Data generation}

To train and test the ML approach, we create a data set by analyzing different resonators through a physics-based FE model. We first define a Latin hypercube sample on the input space, with $\Num \gg 10000$ points. About half of these remain after discarding the designs violating the constraints. We randomly select 11000 points from this set, forming the input data set. 

The designs are loaded into commercial FE software (ANSYS 2022R2 \cite{Ansys2022R2}) which constructs the geometry using the embedded CAD software module, creates a mesh, and runs a preset modal analysis. In the modal analysis, the bottom face of the cylinder is subjected to a fixed boundary condition. The software calculates the twelve lowest-frequency eigenmodes. After the computation, the associated frequencies and modal masses are written to disk. Because of various issues (mesh failures, licensing errors, etc.), certain designs fail to compute. Finally, 10513 complete data pairs containing the input and output vector are obtained. The dataset is available under \url{https://renkulab.io/datasets/3d1db08396c84cb1994598e8349e8a9c}.

\subsection{Network architecture and training}
\begin{figure}[h]
    \centering
    \includegraphics[width=.9\textwidth]{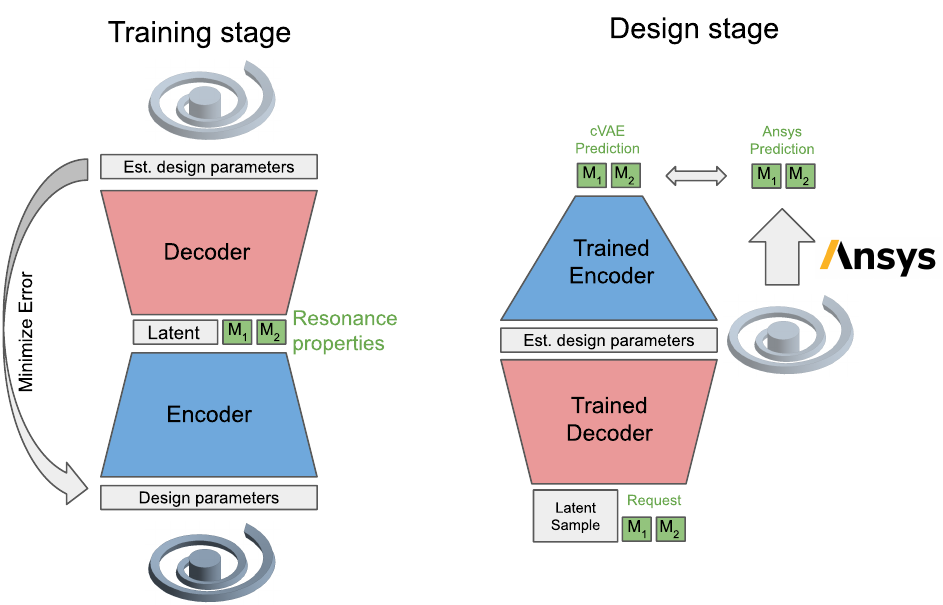}
    \caption{Schematic overview of the cVAE pipeline for resonator design. During the training stage (left) encoder and decoder are trained, such that a latent space is learned allowing for reliable (conditional) reconstruction. During the design stage (right) latent samples and required dynamic properties are provided and the proposed design is then Validated with CAD software.}
    \label{fig:cVAE_scheme}
\end{figure}

\begin{table}[h]
\centering
    \caption{Settings for training the conditional VAE.}
    \label{tab:cvae_settings}
    \begin{tabular}{lll}
        \toprule
        Settings     				            &	Value	  \\
        \midrule
        Latent dimensions    	                &	4         \\
        Encoder input                           & log design parameters \\
        Encoder hidden layers                   &  	[64,64]   \\
        Encoder layer type                      & Residual block   \\
        Encoder loss                            & KL divergence + MSE \\
        Decoder input                           & latent space sample + log performance attributes \\
        Decoder hidden layers      	            &	[64,64]   \\
        Decoder layer type                      & Residual block \\
        Decoder loss                            & MSE \\
        Batch size                   	        &	128       \\
        Optimizer                               &	ADAM      \\
        Validation set split                    &   10 \%        \\
        Convergence criterion                   &	Early stopping (patience=12)\\
        \bottomrule
    \end{tabular}

\end{table}

The cVAE has two main building blocks which are shown in Fig.~\ref{fig:cVAE_scheme}. First, the variational encoder, in blue, uses the design parameters as input values, and its output is the mean and log variance of a Gaussian distribution in the latent space. In addition, a simple surrogate forward model provides an estimate of the requested attributes, namely the frequencies and modal masses of the first $N=2$ eigenmodes, symbolically represented by M$_1$ and M$_2$. The second part, in red, is the decoder taking a point in the latent space and the requested attributes as input, to map these back to the set of design parameters.

Both encoder and decoder are neural networks with two fully connected residual layers, each having $64$ hidden units and leaky rectified linear units (ReLU)~\cite{Xu2015EmpiricalNetwork} as an activation function. Furthermore, we specify the latent space to be four-dimensional. Using the logarithm of the performance attributes for training avoids predicting negative frequencies and masses, and handles the different orders of magnitude in the data. 

To train the network we need to minimize the losses for the different building blocks (Fig~\ref{fig:cVAE_scheme}, left). For the encoder, this is the Kullback-Leibler (KL) divergence between the output Gaussian density and standard normal distribution. An additional loss is the mean squared error (MSE) between the true and the encoder's predicted performance attributes. The loss of the decoder is also the MSE between the design parameters of the training points and its predictions. $80\%$ of the generated designs are used as the training set, $10\%$ are kept as a validation set, and the remaining designs as the test set. In practice, the cVAE is trained with mini-batches of $128$ designs to minimize the total loss until no improvement is observed anymore on the validation training set. 
An overview of the cVAE settings is shown in Table~\ref{tab:cvae_settings}.

For fast on-demand resonator designs (Fig.~\ref{fig:cVAE_scheme}, right) we provide the trained decoder with requested dynamic properties. Random samples from the standard Gaussian density in the latent space (shown in gray) while keeping the requested attributes constant (in green) each yield a different design, so that an arbitrary number of geometries can be generated. For all the designs we predict the dynamic properties via the trained encoder, and take the samples with the lowest design error as the most promising candidates for the requested properties. Their accuracy can then additionally be validated via the FEM software. The cVAE code is available at \url{https://renkulab.io/projects/christian.donner/resonator-design}.

\section{Assessment of the cVAE design accuracy}
\subsection{Analysis of the cVAE prediction error}
\begin{figure}
    \centering
    \includegraphics[width=\textwidth]{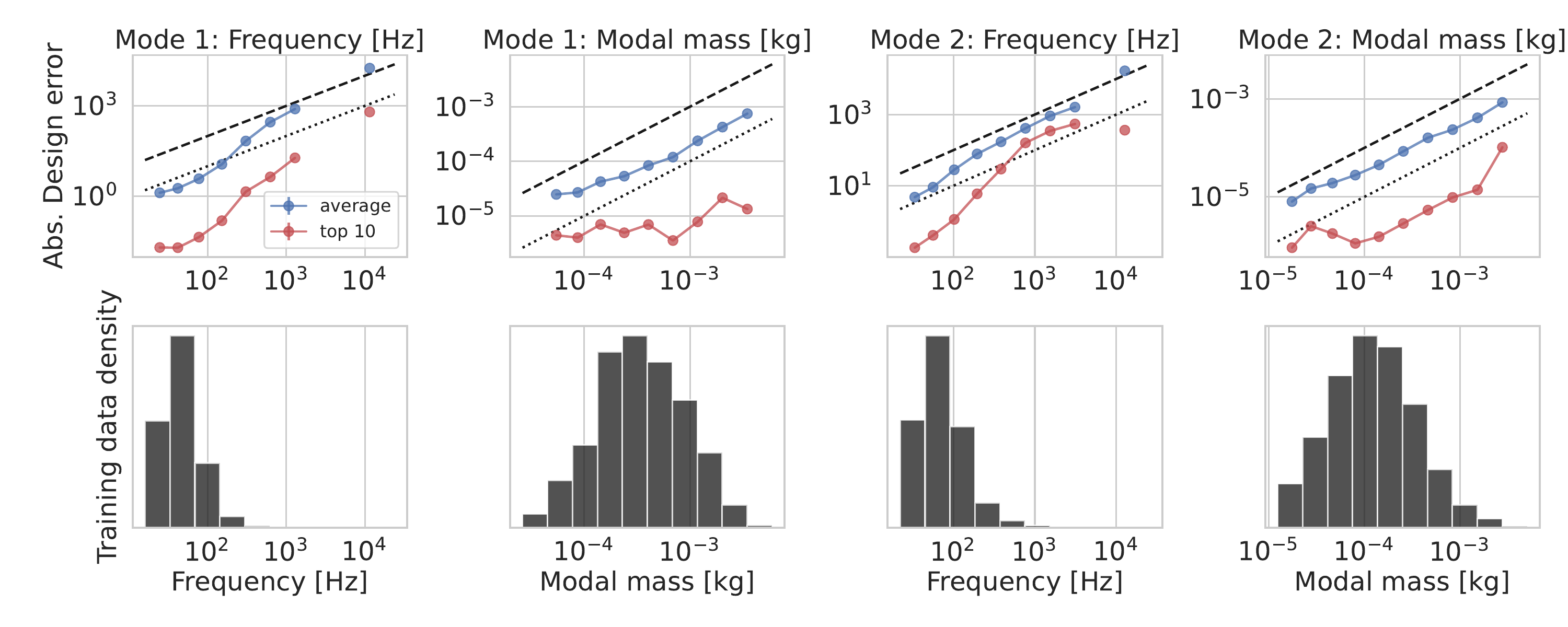}
    \caption{The predicted {\it design errors} for different performance attributes in the top panels. Blue line is the average over $1000$ generated designs and red line indicates the error selecting the 3 designs with lowest error. Dashed line indicates identity (error is on the order of magnitude as request), and dotted line marks the line, where the error is one order of magnitude lower than the request. Lower panels display the histograms of the training data.}
    \label{fig:vae error}
\end{figure}
After having trained the cVAE as described above, we first check the consistency of the model. We investigate if the cVAE accurately predicts performance attributes (the modal properties) for generated designs. To this end, we compute the design error on the modal frequencies and masses of the held-out test set, to ensure that a certain combination of properties can be physically achieved. The absolute design error for feature $i$ is defined as
\begin{equation}
    \text{design error}_i =\vert y^{\rm pred}_i-y^{\rm request}_i\vert,
\end{equation}
where $y^{\rm request}_i, y^{\rm pred}_i$ are the requested attributes for the target value, and the prediction of the cVAE's forward model for the designs, respectively. Figure~\ref{fig:vae error} shows the average design errors for $1000$ generated designs, and for the top 3 designs those with the lowest summed design errors. We observe that the error of the top 3 designs is at least one order of magnitude lower than the requested value, meaning we can expect errors $<10\%$. As expected, errors increase for areas in the design space where we provided less training data, emphasizing the importance of covering the desired design space well with the training data. In the range with sufficient training values, the relative design error drops below 1\%, which is deemed sufficiently accurate for our purposes.

\begin{figure}
    \centering
    \includegraphics[width=\textwidth]{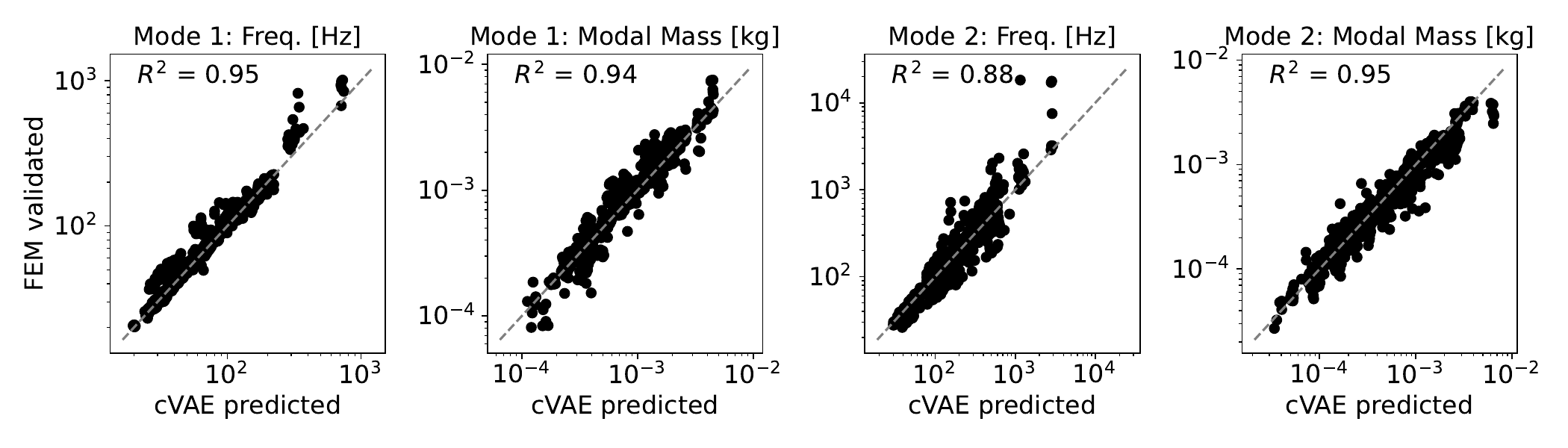}
    \caption{Comparing the predicted performance attributes to FEM validation. On the x-axis are the cVAE predicted performance attributes for a given design and on the y-axis the performance attributes validated by FEM software ANSYS. The $R^2$ score is calculated in log-space.}
    \label{fig:vae vs fem}
\end{figure}

To ensure that the design error is a sensible measure, the cVAE encoder has to predict the performance attributes well. To this end, we validate the cVAE predictions with FEM software (Figure~\ref{fig:vae vs fem}). For 100 test samples we generate 1000 resonators from the cVAE, and take the top 10, i.e. the ones with the lowest design error. Then we evaluate these designs with the FEM software and compare the cVAE-predicted performance attributes with the values obtained through the FEM analysis. Furthermore, we report the $r^2$ value for each performance attribute evaluated in log space. The results show a very good correlation between the cVAE predictions and FE results, meaning that the design selection based on the fast cVAE ranking is defendable. Only the frequencies of the second mode have an $R^2$-value lower than 0.9, but this is mainly due to few training values at higher frequencies.

\subsection{Performance of the cVAE with smaller training sets}
\begin{figure}
    \centering
    \includegraphics[width=\textwidth]{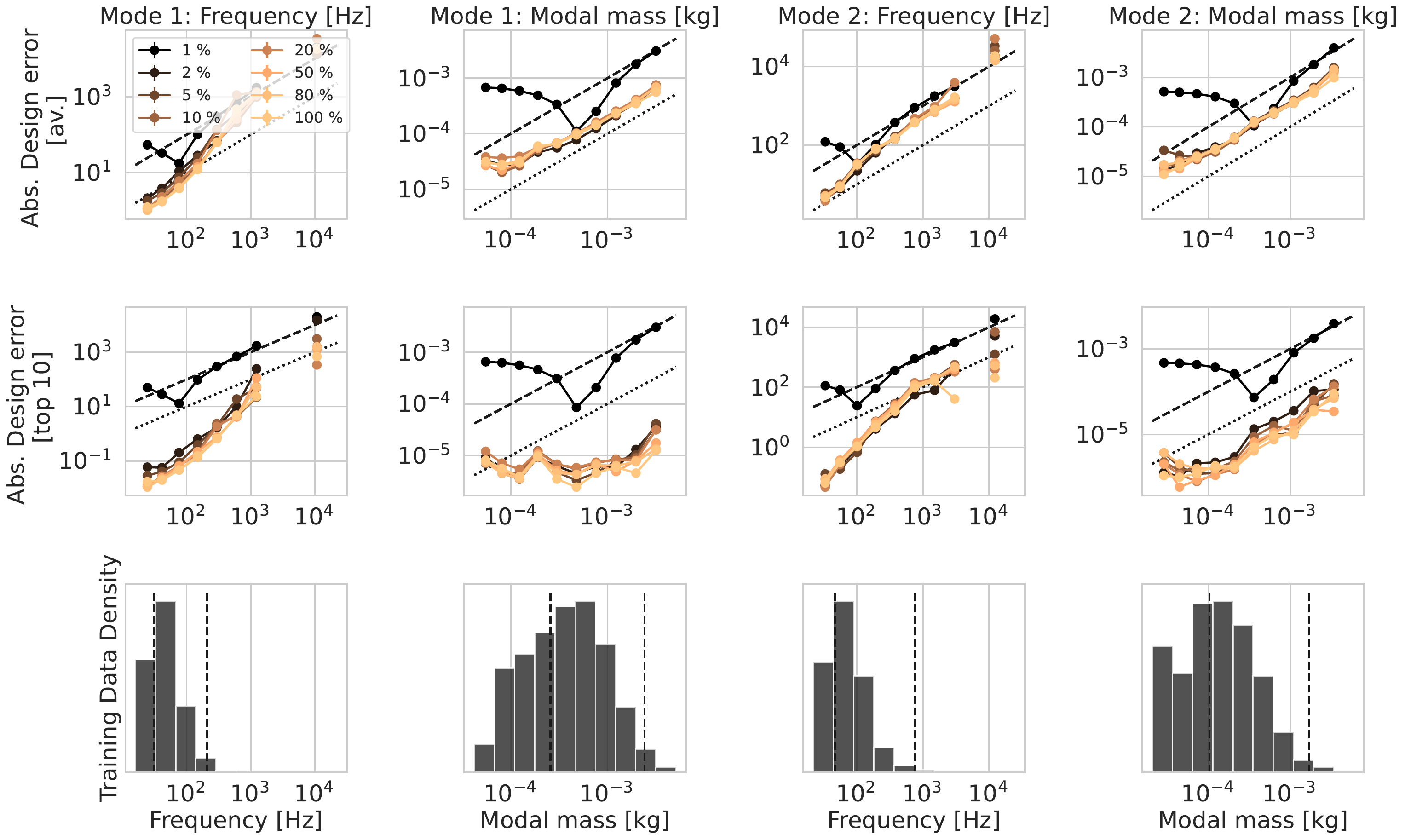}
    \caption{Sensitivity of design errors with respect to the amount of training data. ($1\%:83,\ 5\%:416,\, 20\%:1666,\ 100\%:8332$ designs). The upper panel shows the average design error, the center the design error of the top 10 designs, and below are the histograms of dynamics properties for all $8332$ designs in the full training set.}
    \label{fig:sensitivity}
\end{figure}

We investigate to what degree the performance of the cVAE depends on the size of the training set. The full data set still requires the generation of 10000 design points for 6 geometry parameters and 4 modal properties, which might not be smaller than the amount of iterations required for classical topology optimization. The cVAE quality assessment in the previous section used the full training set with $>8000$ designs, and this approach is repeated for increasingly small random subsets of the training data. Figure~\ref{fig:sensitivity} shows that the design error attains similar levels when using as little as $5\%$ (416) designs of the training data. This is a strong indicator, that only a fraction of the used data is needed for training a cVAE to yields accurate designs. However, a thorough analysis of the training data needed for a given size of input and output vectors is outside the scope of this work.

\subsection{FEM validation of cVAE designs in comparison to training-set look up}
Next, instead of relying on the cVAE prediction of the performance attributes, we use ANSYS FE software for validating the cVAE method on a dataset of 100 test designs. We first train the cVAE on $5\%$, or $20\%$ of training samples (416, 1666 designs respectively). Then for each test design, we generate 1000 surrogate designs with the cVAE, from which we keep the top 10 designs where the cVAE predicts the performance attributes that are closest to the requested ones. Till this point, the generation is fast and cheap. For the remaining 10 geometries we calculate the true performance attributes with the FE software. Finally, we keep the design that yields the performance attributes closest to the requested ones. 

\begin{figure}
    \centering
    \includegraphics[width=.8\textwidth]{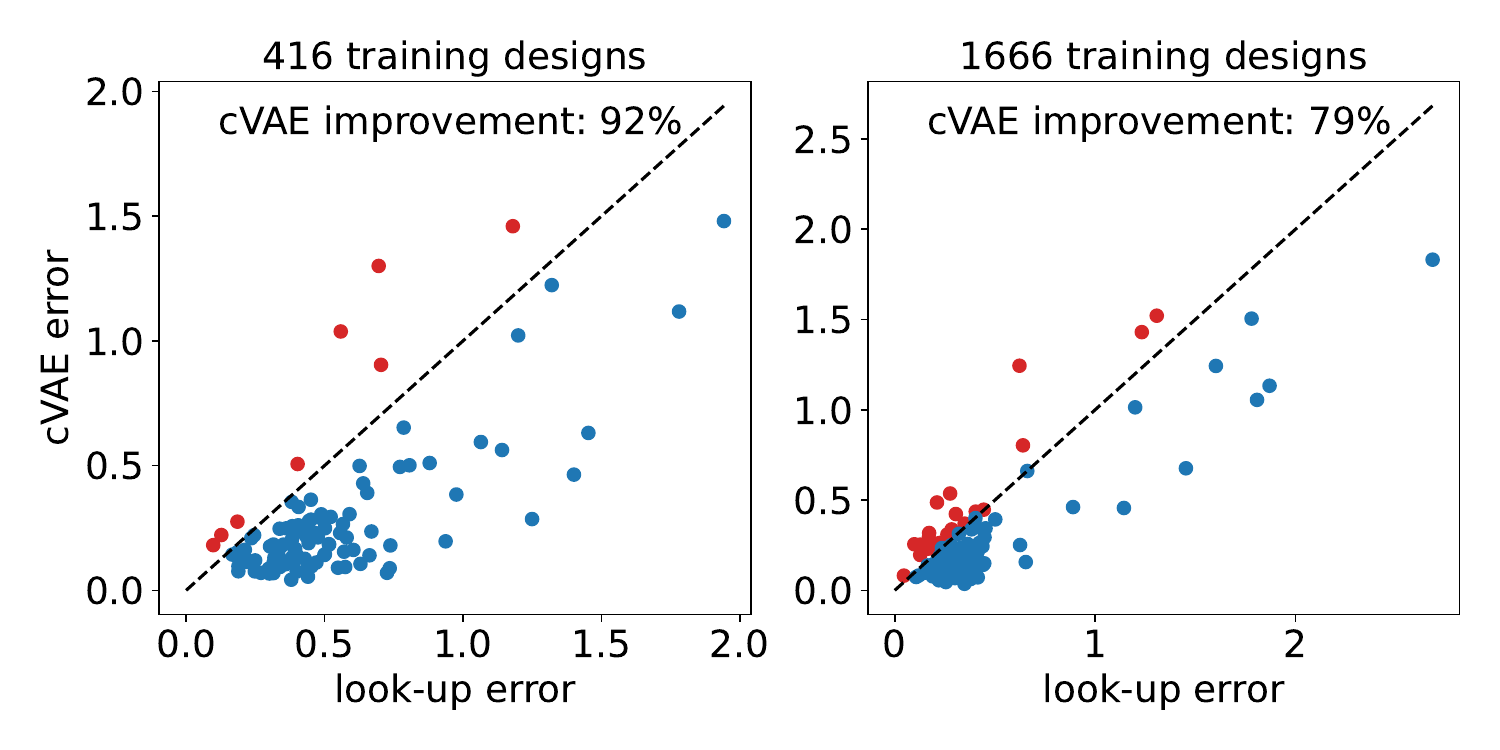}
    \caption{Comparing the cVAE performance against design look-up from the training set. Left and right panels with 416, 1666 spiral designs were used for cVAE training and look-up, respectively. The performance was evaluated for 100 new designs. Each dot depicts the mean design error for the look-up method vs cVAE method. Blue dots denote designs, where the cVAE error was lower, and red dots, where the look-up error was lower. The cVAE provided an improvement in $92\%,78\%$ over the look-up method, respectively.}
    \label{fig:look-up}
\end{figure}

The strength of the generative resonator design can finally be illustrated by comparing it to a look-up strategy as a baseline. This method provides the training design as the prediction with the smallest distance to the requested performance attribute. As a distance measure, the Euclidean distance between the logarithmic performance attributes is used to account for the fact that modal masses and frequencies account equally. The results are shown in Figure~\ref{fig:look-up}. We see, that the cVAE designs yield the smaller overall design error in 92\% (416 training designs) or 79\% (1666 training designs) of the cases. 

These results support the chosen method as a very efficient tool in resonator design: for 6 geometric input parameters and 4 attributes, a training set containing only 400 modal analyses is sufficient to realize an intricate optimization. Although the prediction error is still low, the quality of the predictions compared to using a lookup algorithm in a very sparsely populated design space is evident. 

\section{Application of cVAE generative design for LRMs with multiple band gaps}
\subsection{Design of resonators with desired properties}
\begin{figure}
    \centering
    \includegraphics[width=\textwidth]{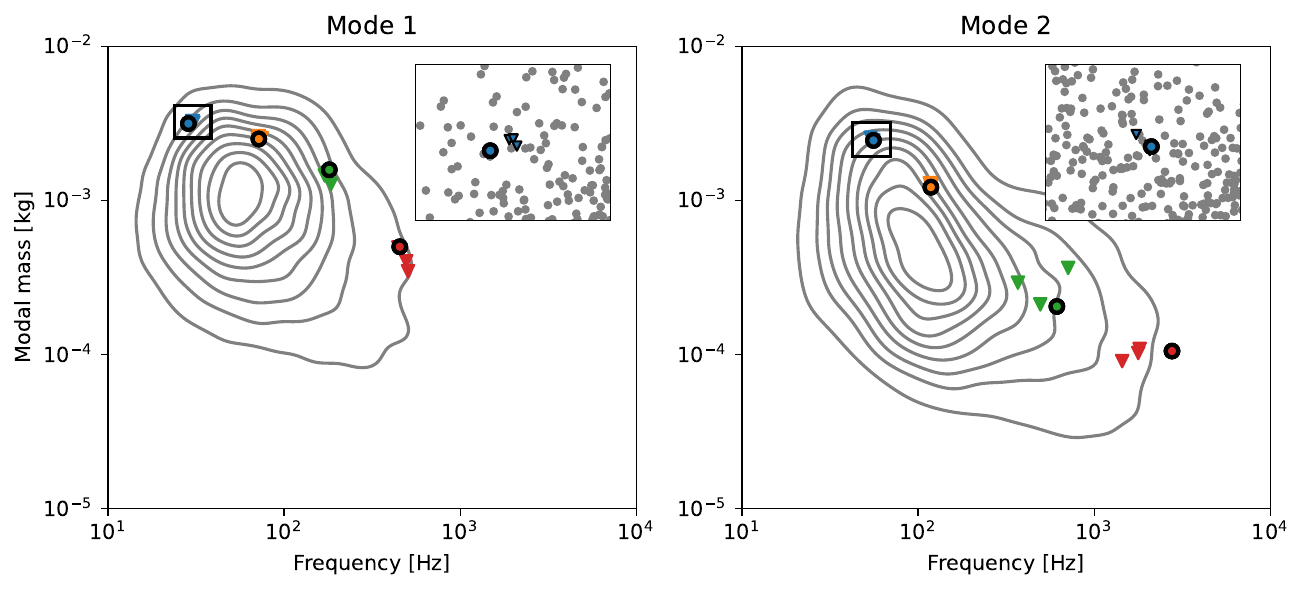}
    \caption{Requested attributes (circles) versus attributes of generated designs (triangles), calculated in Ansys for the 20 best-ranked designs. The different colors indicate the four different requests. The contours show the distribution of the full cVAE training data set. The insets show the enlarged area of design A. Gray dots are the training designs.}
    \label{fig:ansys}
\end{figure}

Having validated the quality of the cVAE predictions, we wish to assess its general applicability by demanding a set of modal parameters which are not included in the generated data set. There is, however, a limitation to the freedom of choice of these parameters since not all requests are physically realizable. Specifying the frequency and mass of mode 1 limits the possible ranges for mode 2 since there is a certain correlation between the modes. Here, we randomly specify the parameters for mode 1, and conditioned on these specifications we sample a multivariate Gaussian distribution to find the range of possible values for mode 2. The Gaussian was previously fitted to the data the cVAE was trained on.

With this restraint, we ask the cVAE $4$ different requests which were not within the generated dataset and let it generate $1000$ designs by sampling the latent space. The top $20$, as predicted by the cVAE, are explicitly modeled in ANSYS to check whether these designs match the requested dynamic properties. Figure~\ref{fig:ansys} visualizes the dynamic properties of the best $3$ designs, showing a good agreement. As expected,  the results become less accurate when coming to the tails of the training distribution, as was shown in Figure~\ref{fig:vae error}. We observe that the requested attributes for mode 1 are matched more closely than those of the second mode. 

\begin{figure}[h]
    \centering
     \begin{subfigure}[b]{0.24\textwidth}
        \centering
        \includegraphics[trim=300 100 300 100,clip,width=\textwidth]{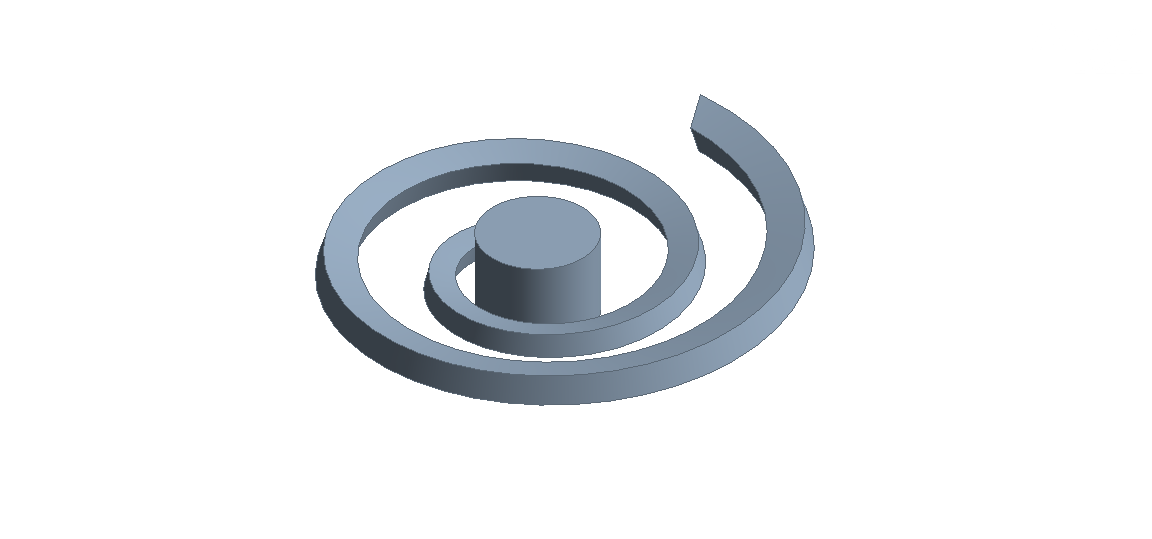}
        \caption{}
        \label{fig:desa}
     \end{subfigure}
     \hfill
     \begin{subfigure}[b]{0.24\textwidth}
         \centering
         \includegraphics[trim=300 100 300 100,clip,width=\textwidth]{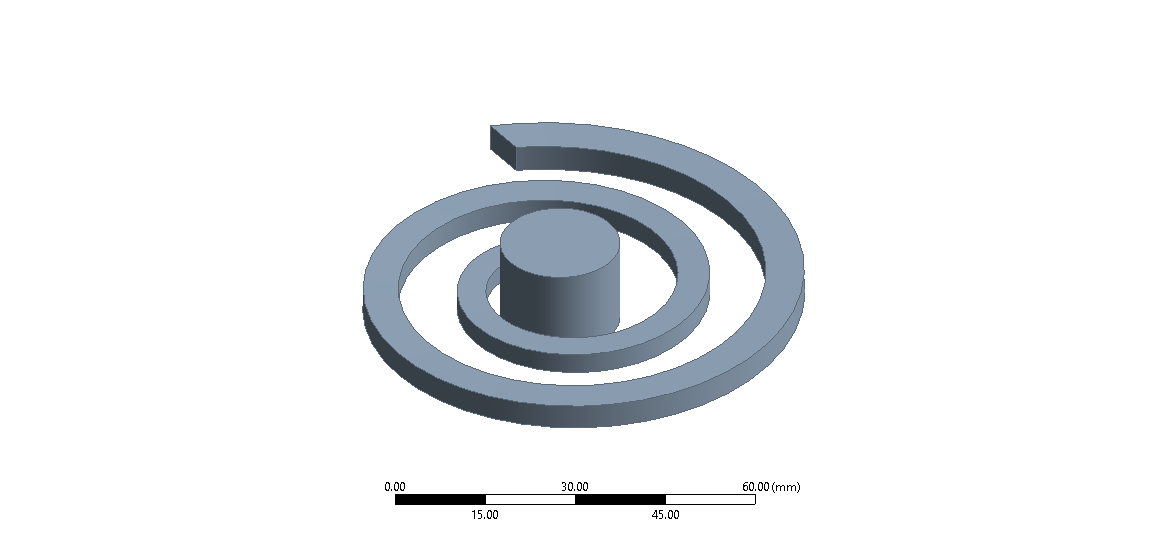}
         \caption{}
        \label{fig:desb}
     \end{subfigure}
     \hfill
     \begin{subfigure}[b]{0.24\textwidth}
        \centering
        \includegraphics[trim=300 100 300 100,clip,width=\textwidth]{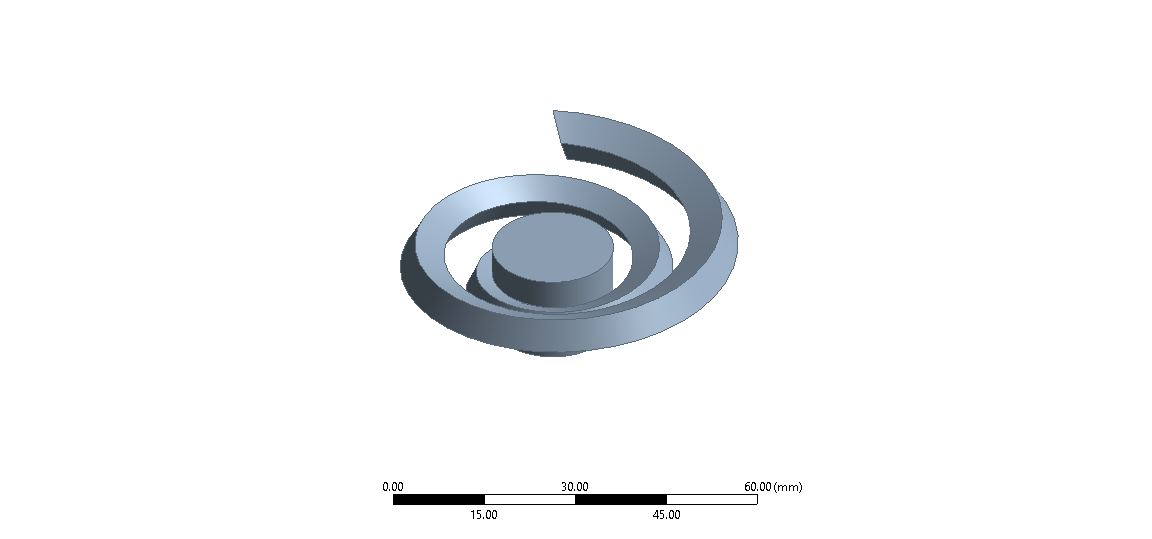}
        \caption{}
        \label{fig:desc}
     \end{subfigure}
     \hfill
     \begin{subfigure}[b]{0.24\textwidth}
         \centering
         \includegraphics[trim=300 100 300 100,clip,width=\textwidth]{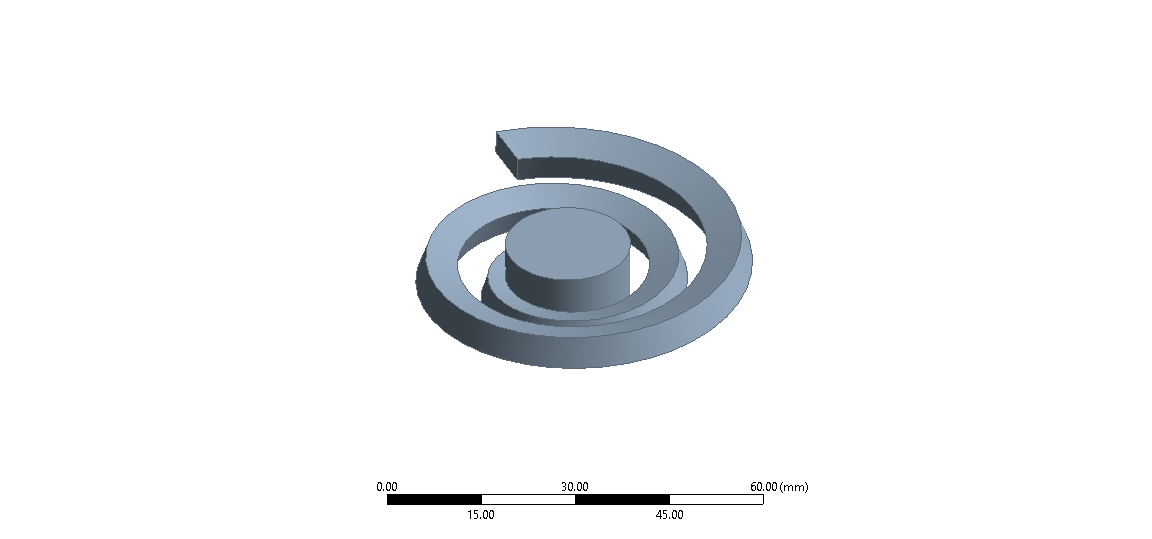}
         \caption{}
        \label{fig:desd}
     \end{subfigure}
     \caption{Resonator designs studied in detail. (a)-(b) Designs A and B, the two best designs for the first request. (c)-(d) Designs C and D, the two best designs for the second request.}
     \label{fig:des}
\end{figure}

\begin{figure}[h]
    \centering
     \begin{subfigure}[b]{0.24\textwidth}
        \centering
        \includegraphics[trim=250 50 250 100,clip,width=\textwidth]{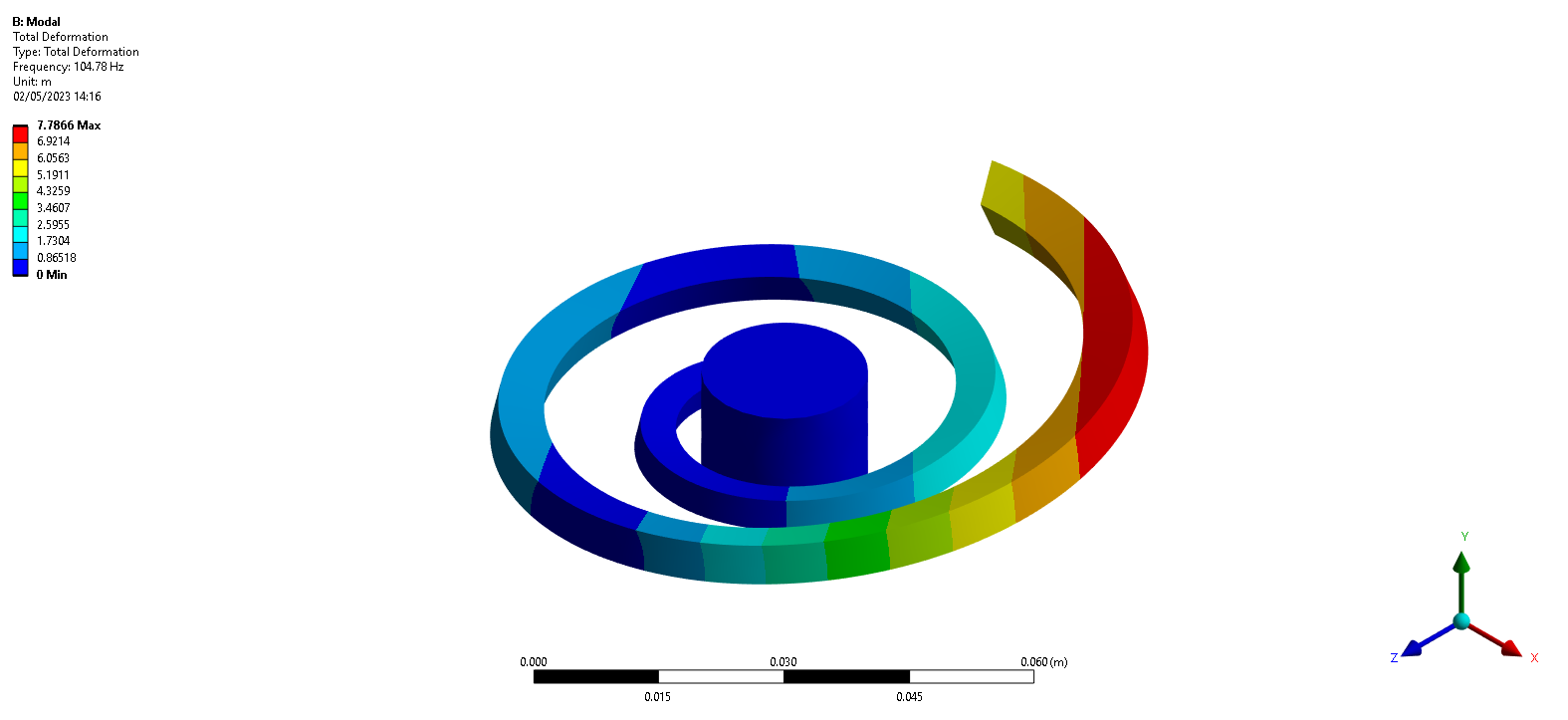}
     \end{subfigure}
     \hfill
     \begin{subfigure}[b]{0.24\textwidth}
         \centering
         \includegraphics[trim=250 50 250 100,clip,width=\textwidth]{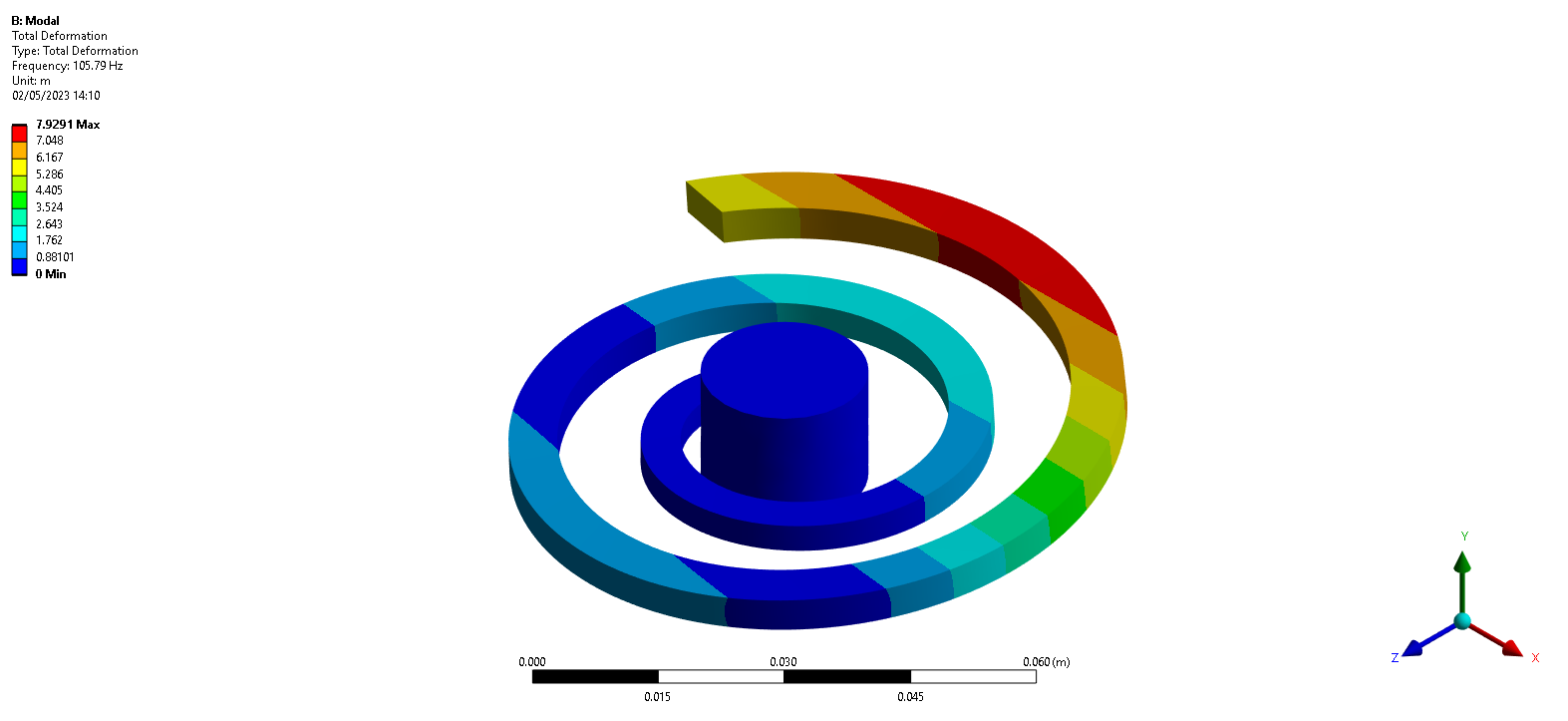}
     \end{subfigure}
     \hfill
     \begin{subfigure}[b]{0.24\textwidth}
        \centering
        \includegraphics[trim=250 50 250 100,clip,width=\textwidth]{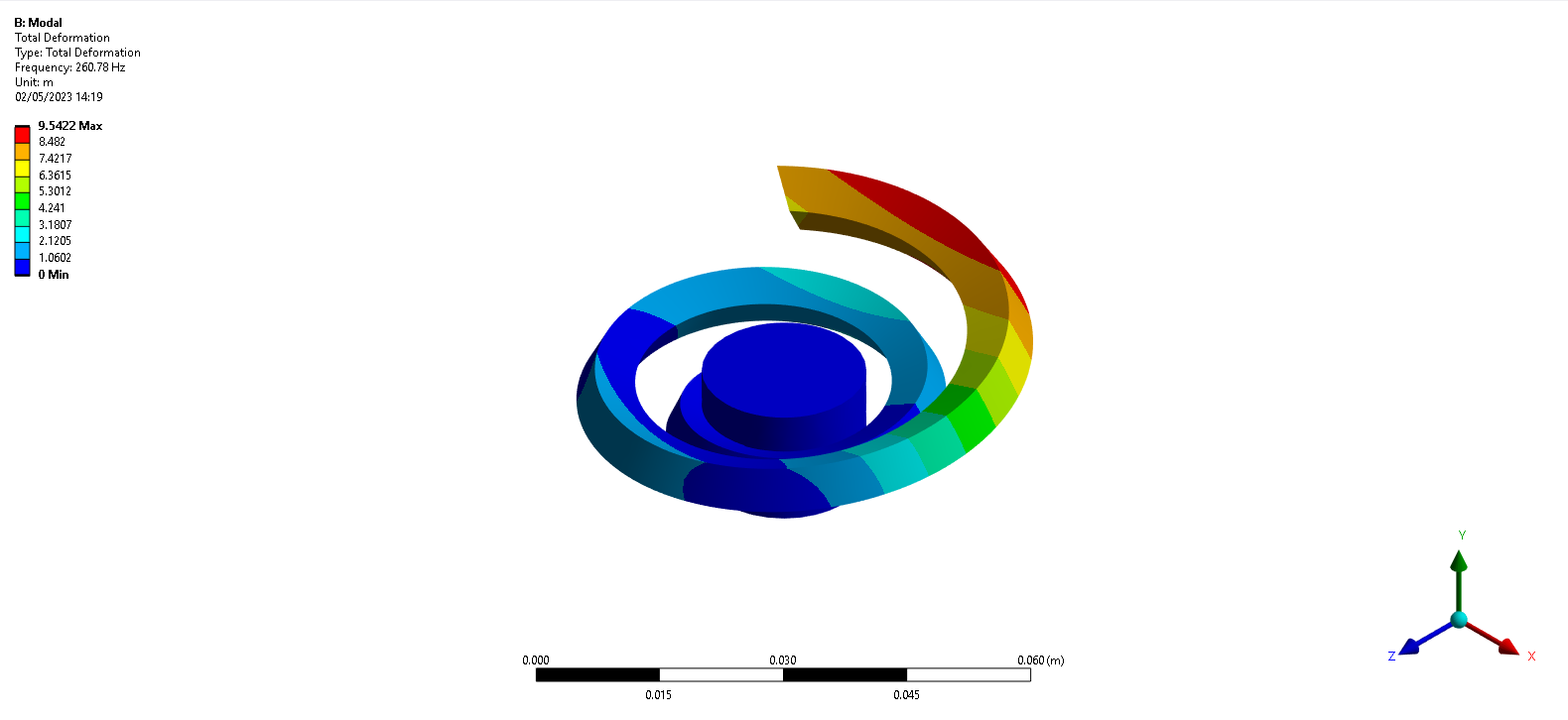}
     \end{subfigure}
     \hfill
     \begin{subfigure}[b]{0.24\textwidth}
         \centering
         \includegraphics[trim=250 50 250 100,clip,width=\textwidth]{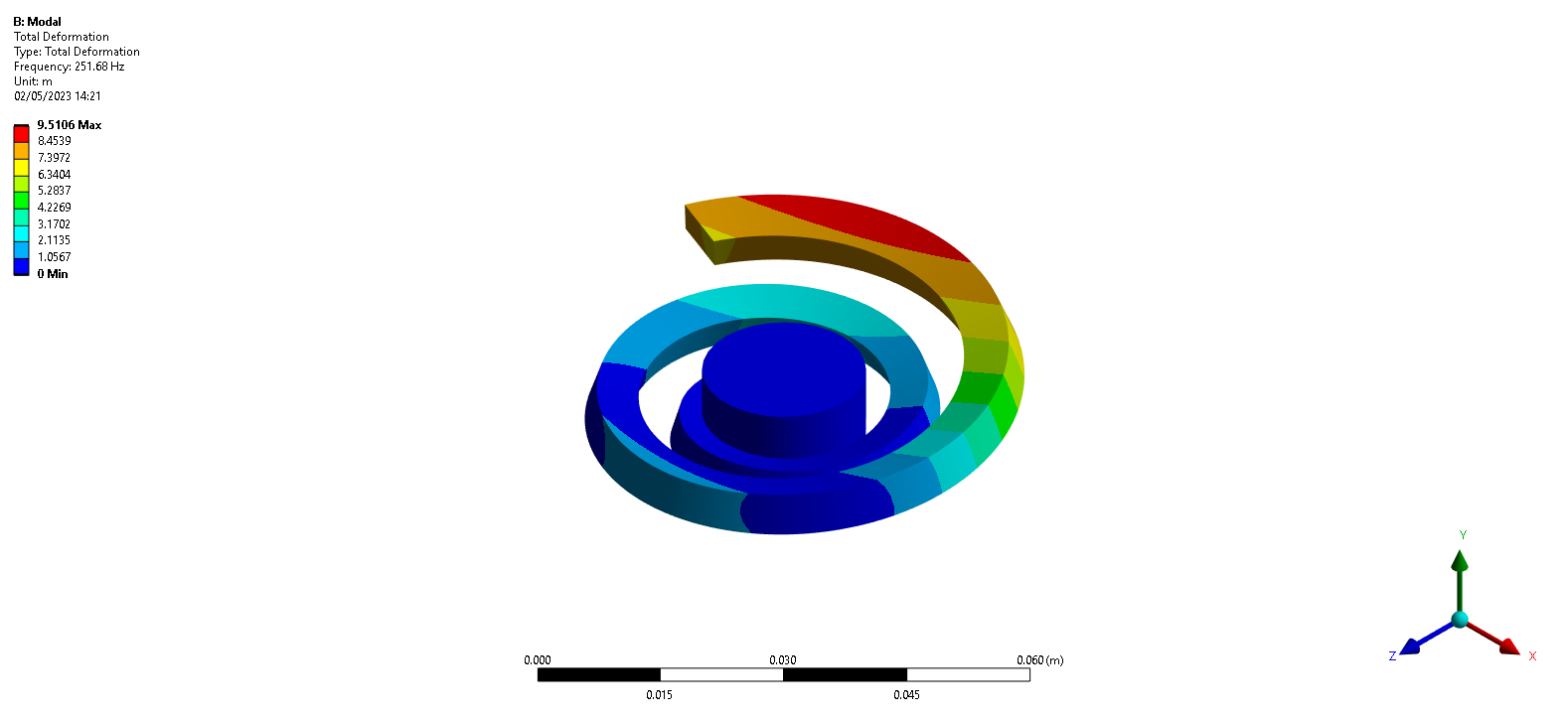}
     \end{subfigure}

     \begin{subfigure}[b]{0.24\textwidth}
        \centering
        \includegraphics[trim=250 50 250 100,clip,width=\textwidth]{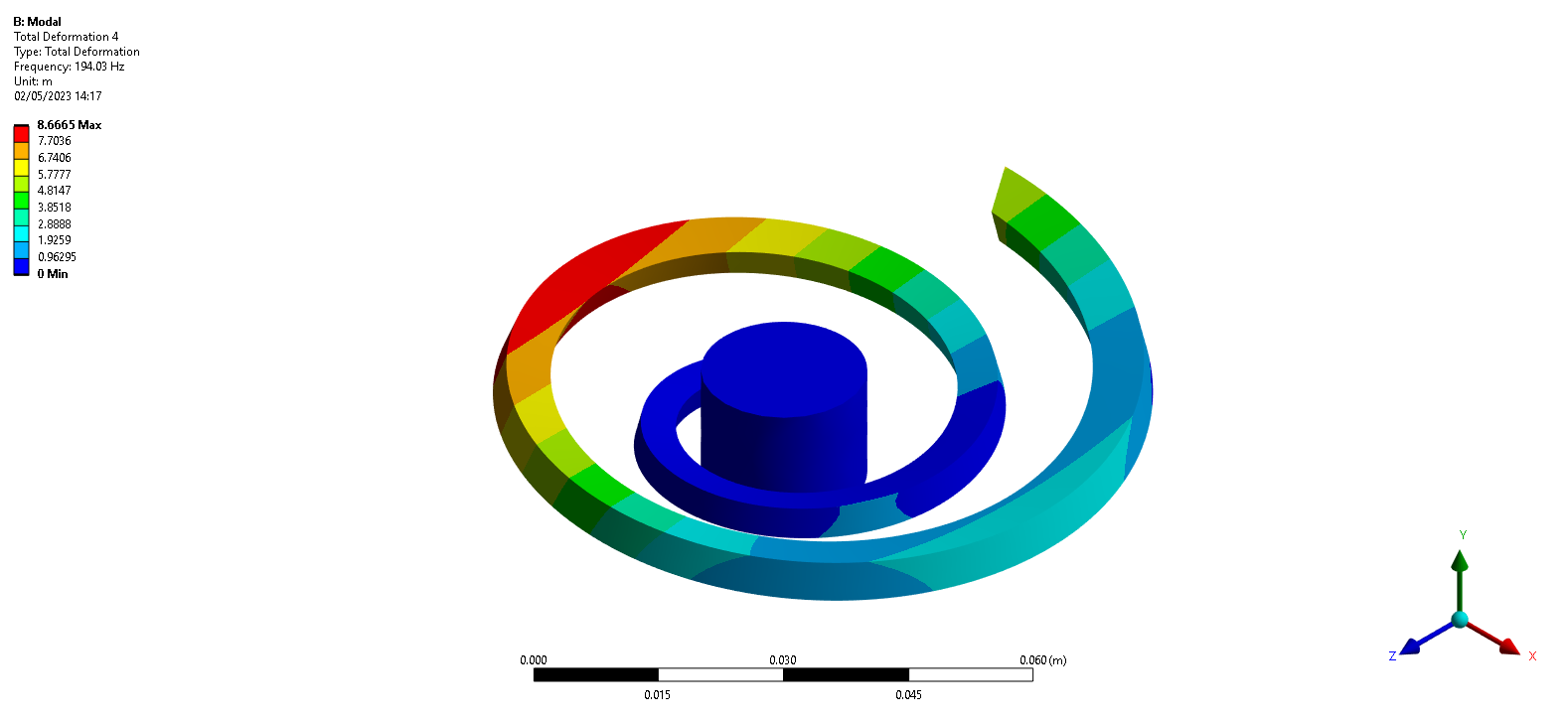}
        \caption{}
        \label{fig:desamod}
     \end{subfigure}
     \hfill
     \begin{subfigure}[b]{0.24\textwidth}
         \centering
         \includegraphics[trim=250 50 250 100,clip,width=\textwidth]{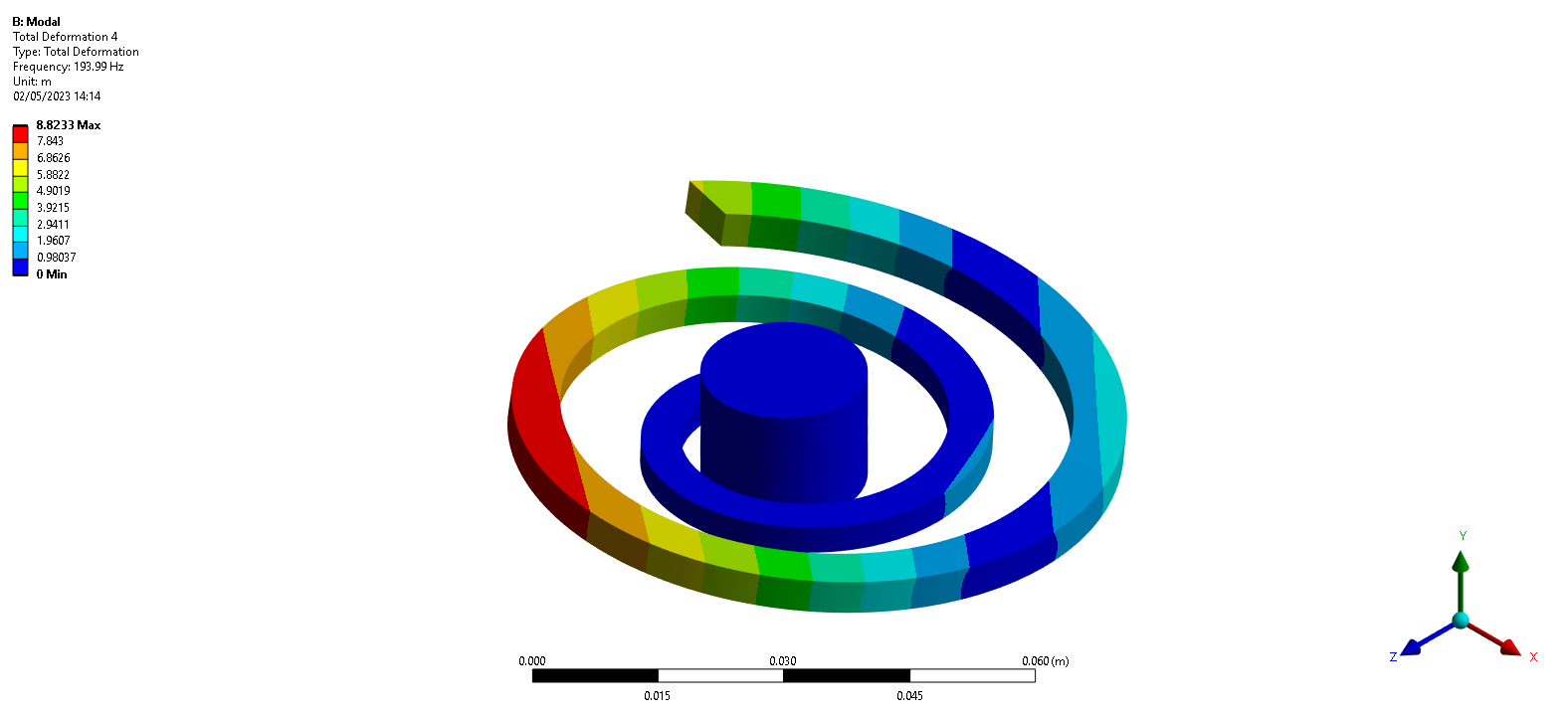}
         \caption{}
        \label{fig:desbmod}
     \end{subfigure}
     \hfill
     \begin{subfigure}[b]{0.24\textwidth}
        \centering
        \includegraphics[trim=250 50 250 100,clip,width=\textwidth]{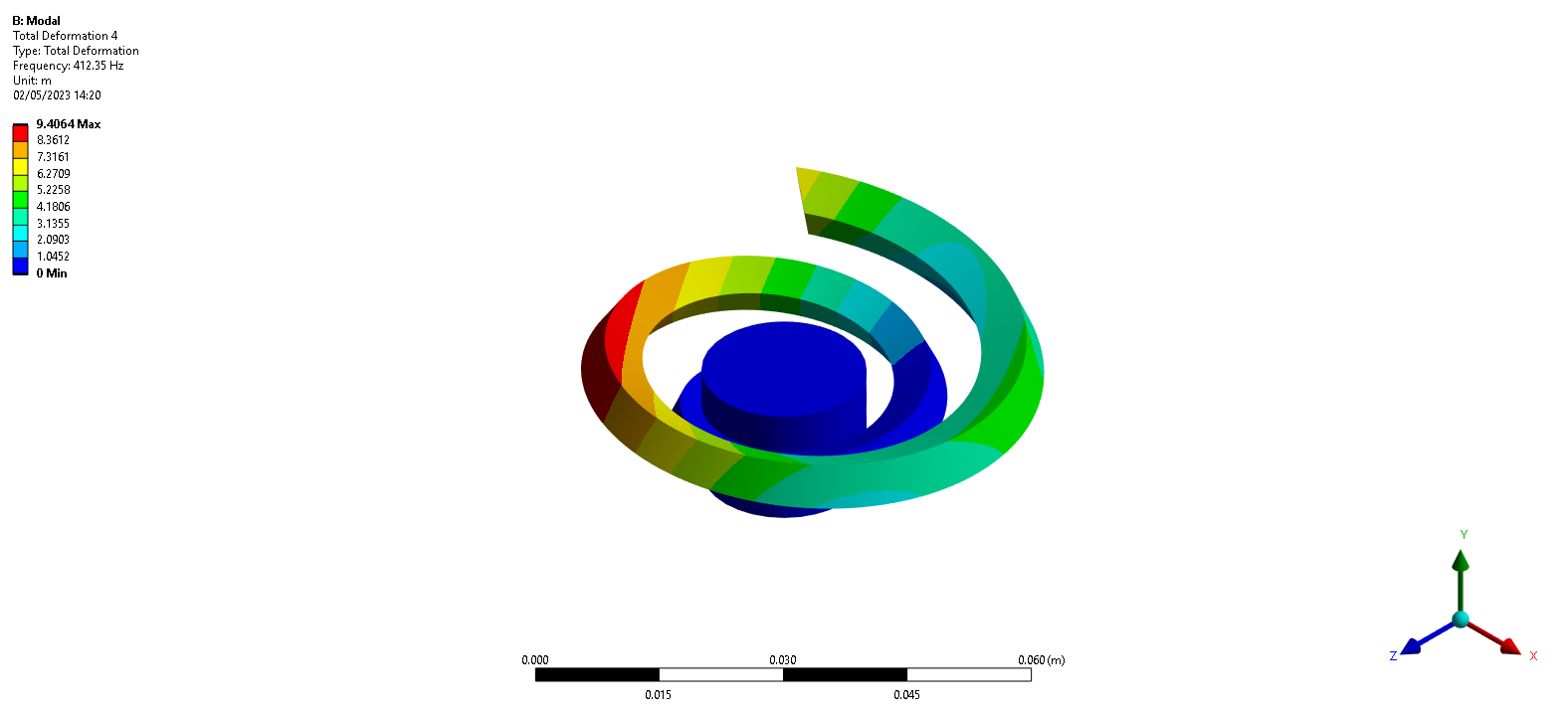}
        \caption{}
        \label{fig:descmod}
     \end{subfigure}
     \hfill
     \begin{subfigure}[b]{0.24\textwidth}
         \centering
         \includegraphics[trim=250 50 250 100,clip,width=\textwidth]{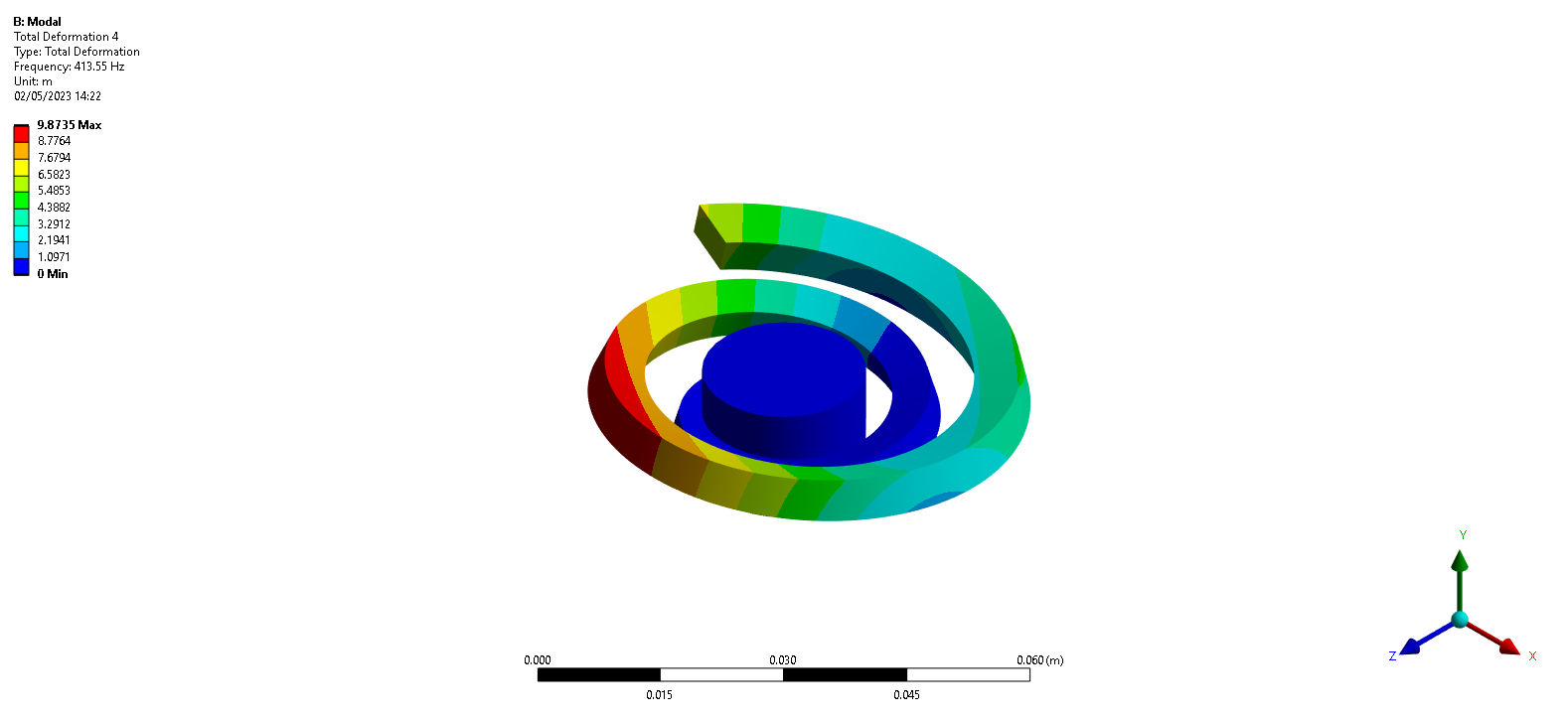}
         \caption{}
        \label{fig:desdmod}
     \end{subfigure}
     
     \caption{Simulated mode shapes for the generated resonator designs. The top and bottom rows show the first and second strongest modes in terms of vertical modal mass. Subfigures (a)-(d) show the four designs A-D.}
     \label{fig:desmod}
\end{figure}

Figure \ref{fig:des} and table \ref{tab:des} show the geometry and modal properties of the top-performing designs. Designs A-B (Figures \ref{fig:desa}-\ref{fig:desb}) and designs C-D (Figures \ref{fig:desc}-\ref{fig:desd}) are the cVAE's best proposals to satisfy, respectively, a request with lower and higher resonance frequencies. It should not be a surprise that asking lower frequencies leads to longer, more slender spirals. However, some design freedom remains, in particular for the amount of twist of the spiral arm. The cVAE appears to be effective in exploiting this, proposing geometrically distinct designs with highly similar modal parameters. In Figure \ref{fig:desmod}, the mode shapes associated with the high-mass modes are depicted. It appears that the ranking in terms of modal mass tends to select the same mode `types' across designs. It is also interesting to note that the spiral shape exhibits tonotopy -- where different frequencies activate different spatial regions -- consistent with the findings of e.g. Dal Pogetto et al. \cite{DalPoggetto2023}

As seen in table \ref{tab:des}, the cVAE designed resonators all give accurate approximations of the target parameters. This holds for both predicted, simulated, and measured values. The accumulation of errors between these estimates remains very limited. Most importantly, the differences arising within the numerical procedures (cVAE prediction, FE simulation) are comparable to the differences between the simulated and measured values. Therefore, in terms of accuracy, the deficit of the cVAE generation is similar to the one inherent to the manufacturing process. 

\begin{table}[]
\centering
\caption{Modal parameters for top-performing designs given two separate requests (lower and higher desired frequencies). For each design, the table lists the target parameters, the cVAE prediction, the FE simulation, and the average measured values.}
\label{tab:des}
\begin{tabular}{llllll}
\toprule
            &            & \textbf{Mode 1}  &                  & \textbf{Mode 2}             &   \\
            &            & Frequency [Hz] & Mass [g] & Frequency [Hz] & Mass [g] \\
\midrule
Design A    & Target     & 28.6          & 3.16             & 55.7          & 2.45            \\
            & Pred.      & 31.0          & 2.77             & 55.8          & 2.41            \\
            & Simul.     & 30.0          & 3.27             & 55.4          & 2.47            \\
            & Meas.      & 29.5          & 3.62             & 54.3          & 2.28            \\
\midrule
Design B    & Target     & 28.6          & 3.16             & 55.7          & 2.45            \\
            & Pred.      & 28.7          & 2.87             & 55.4          & 2.41            \\
            & Simul.     & 30.2          & 3.28             & 55.4          & 2.42            \\
            & Meas.      & 30.3          & 3.72             & 55.6          & 2.32            \\
\midrule
Design C    & Target     & 71.8          & 2.51             & 117.9         & 1.22            \\
            & Pred.      & 73.4          & 2.54             & 120.0         & 1.19            \\
            & Simul.     & 74.5          & 2.54             & 117.9         & 1.23            \\
            & Meas.      & 73.9          & 2.68             & 117.6         & 1.23            \\
\midrule
Design D    & Target     & 71.8          & 2.51             & 117.9         & 1.22            \\
            & Pred.      & 72.3          & 2.58             & 116.8         & 1.18            \\
            & Simul.     & 71.9          & 2.61             & 118.2         & 1.27            \\
            & Meas.      & 71.5          & 2.66             & 117.1         & 1.23            \\
\bottomrule
\end{tabular}
\end{table}

\subsection{Experimental techniques used for model validation}
\label{ssec:meas}

To complete the validation of the data-driven method, we physically realize the best designs proposed by the cVAE to perform measurements. Two types of tests are executed in the present study. First, in order to estimate the modal properties of the design prototypes, the mechanical impedance is measured directly. The sample is fixed onto an impedance head (PCB Piezotronics 288D01), which connects to a solid brass block that acts as a mechanical high-pass filter. The assembly is excited by a shaker (Brüel \& Kj\ae r LDS V201). The experimental modal properties are found by matching the dynamic mass expression in eq.~\eqref{eq:dynmss} to the measurement via a parametric fit. We use MATLAB's non-linear least-squares routine, where the starting values of each parameter $m_i$ and $\nu_i$ are provided by the FE model. The resulting modal parameters are given in Table~\ref{tab:des}.

\begin{figure}[h]
    \centering
        \centering
        \includegraphics[trim=300 100 300 1000,clip,width=\textwidth]{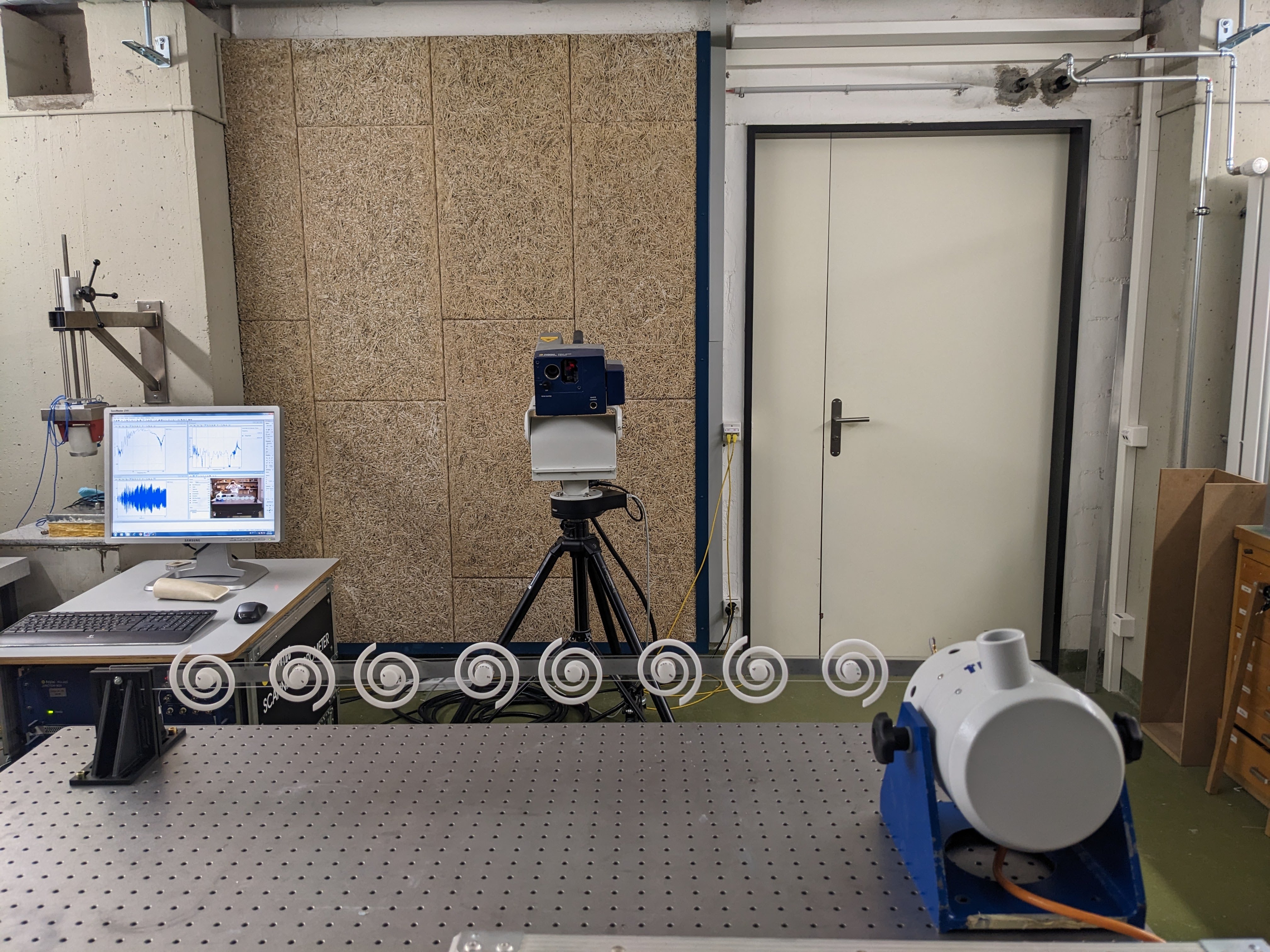}
     \caption{Setup used to measure the dispersion characteristics of the metamaterial beams.}
     \label{fig:meas}
\end{figure}

The second type of test aims to identify the band gaps that the resonators induce in a real metamaterial beam. As a host structure, we consider a beam made of acrylic glass (PMMA), with thickness 3~mm, width 50~mm, and length 1000~mm. The material is characterized by a Young's modulus of 4.8\,GPa, density of 1190\,kgm$^{-3}$, Poisson's ratio of 0.35, and structural damping coefficient of 0.06 \cite{Dedoncker2022a}. For a given design, eight resonators are glued onto the host at 110\,mm intervals. To run the test, the beam is clamped on one end and excited by a shaker (TIRA TV 51120) on the other. We measure the input force (PCB Piezotronics 208C01) and the resulting velocity along the beam (Polytec PSV-400 laser scanning vibrometer). This full-field vibration measurement is finally converted to dispersion curves by the inhomogeneous wave correlation (IWC) method \cite{Berthaut2005,VanBelle2017,VanDamme2018}. The full test setup is depicted in Figure \ref{fig:meas}. 

\subsection{Performance of the LRM with predefined band gaps}
The target frequencies and modal masses were chosen so that the resonators yield an LRM with two distinct band gaps. Thanks to the good approximation of modal parameters for all four designs, the generated resonators have the potential to produce band gaps that very accurately match those of the target. Since the maximum wave number in the considered frequency range is less than 28~rad/m, the minimum wave length is larger than 228~mm. Therefore, the LRM criterion that the resonators should be closer than half a wave length apart is valid in this case.

\begin{figure}[h!]
    \centering
     \begin{subfigure}[b]{\textwidth}
        \centering
        \includegraphics[width=\textwidth]{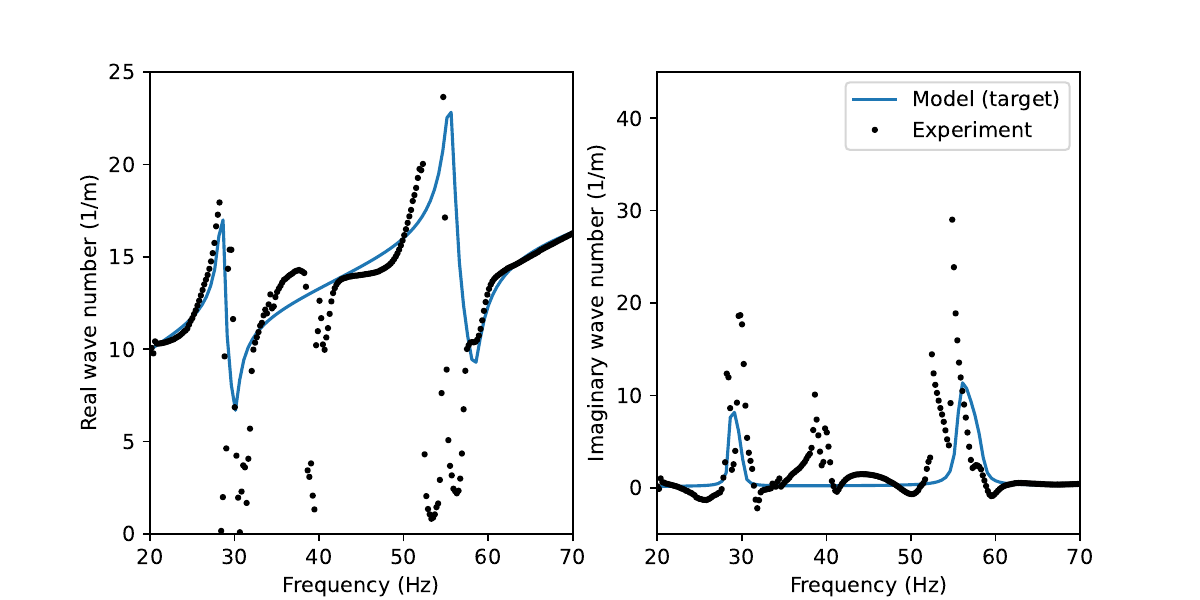}
        \caption{}
        \label{fig:dca}
     \end{subfigure}
     \begin{subfigure}[b]{\textwidth}
         \centering
         \includegraphics[width=\textwidth]{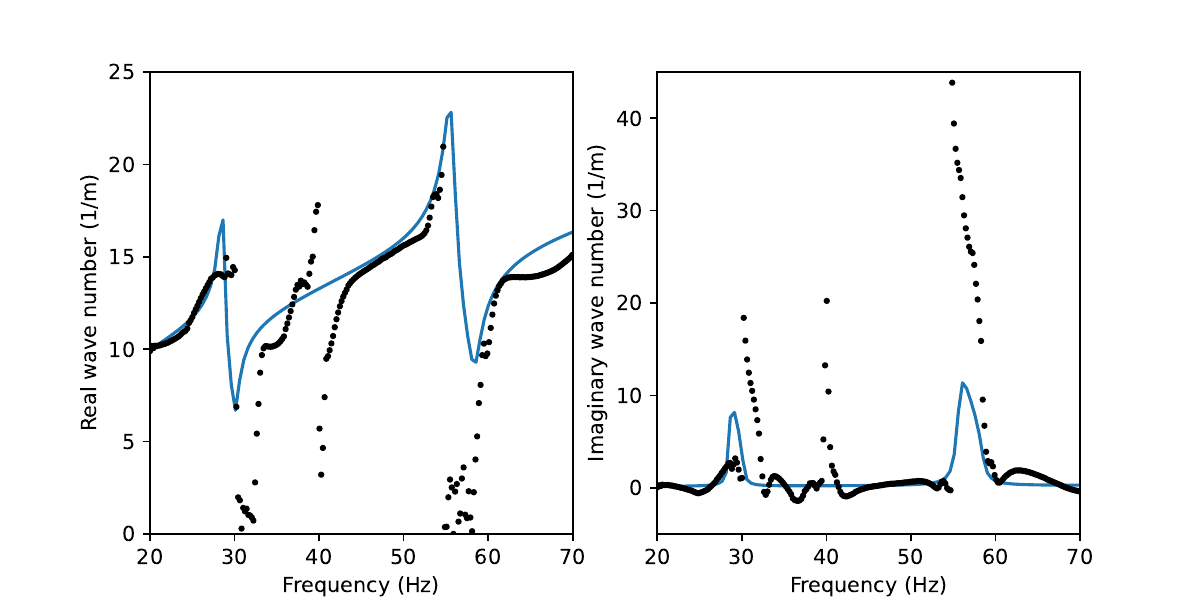}
         \caption{}
        \label{fig:dcb}
     \end{subfigure}
     \caption{Dispersion curves (real and imaginary parts) for a LRM beam with the automatically generated resonator designs. Solid lines result from a model using the target modal parameters (table \ref{tab:des}). Dots indicate the measured wave numbers. (a) Design A (b) Design B.}
     \label{fig:dc1}
\end{figure}

\begin{figure}[h!]
    \centering
     \begin{subfigure}[b]{\textwidth}
        \centering
        \includegraphics[width=\textwidth]{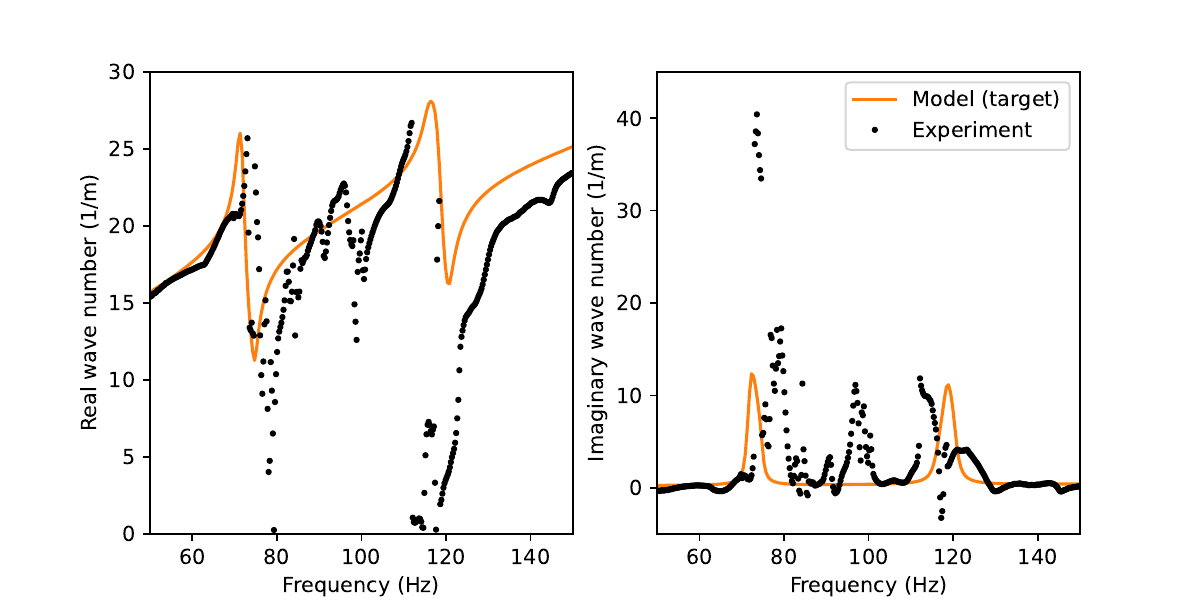}
        \caption{}
        \label{fig:dcc}
     \end{subfigure}
     \begin{subfigure}[b]{\textwidth}
         \centering
         \includegraphics[width=\textwidth]{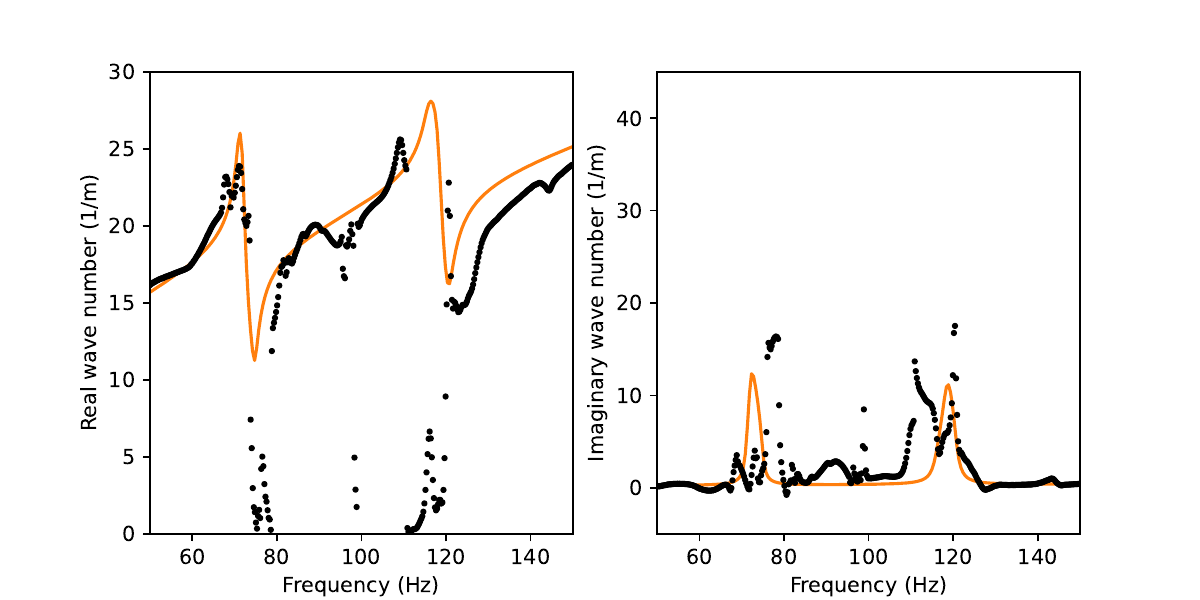}
         \caption{}
        \label{fig:dcd}
     \end{subfigure}
     \caption{Dispersion curves (real and imaginary parts) for a LRM beam with the generated resonator designs. Solid lines result from a model using the target modal parameters (table \ref{tab:des}). Dots indicate the measured wave numbers. (a) Design C (b) Design D.}
     \label{fig:dc2}
\end{figure}

The target and measured dispersion are shown in Figures \ref{fig:dc1} (resonators A and B) and \ref{fig:dc2} (resonators C and D). The correspondence between the modeled (target) and measured curves is good. The location and width of the two target band gaps are accurately predicted by including the two dominant resonator modes in the model. Peaks can appear more spread out, probably due to the slight physical differences between the resonator samples and deviations in material damping due to the sintering process. 

However, as expected, the measured dynamics are richer than the modeled behavior. In all cases, we measure a weaker, third band gap between the main gaps. Upon closer inspection, this is explained by the presence of an additional resonator mode, with an intermediate frequency but a lower modal mass. Since the modes were selected by decreasing modal mass, this third mode is not considered in the cVAE design.

\section{Conclusions and future work} 
In this work, we propose and test a novel but practical targeted design approach for LRMs based on ML techniques. Three-dimensional resonator designs can be evaluated mainly in terms of a few scalar modal parameters, which yield multiple band gaps at desired frequencies and with predefined efficiency. Their modal parameters can be used as performance attributes in a cVAE neural network, an innovative AI solution in the field of structural dynamics. It combines the strength of random sampling of the latent space with the additional conditioning by the requested modal properties. Consistency with the physical properties of the data is thus ensured. After an offline training process -- relying on numerical modal analyses -- we obtain an ultra-fast method for resonator and LRM design. There are two main advantages. The first is that multiple modal properties (leading to multiple band gaps) can be requested simultaneously. The second is the inherent possibility to generate many designs, out of which the most practical one can be chosen.

The approach is validated by applying it to a family of spiral-shaped resonators, determined by six scalar parameters which form the design space. Around 10000 random designs are sampled and then analyzed using finite element simulations, yielding the two modes with the largest dynamic mass for each design. A cVAE with relatively compact encoder and decoder networks is trained on this dataset. The dependencies between in- and outputs are shown to be captured very well in this manner. The cVAE predictions of modal parameters are in line with the `true' values. The decoder part of the cVAE proves capable of generating several different high-quality designs that match the target attributes exceptionally well, deviating less than 5\% from the true modal properties. Even at the outer edges of the design space, where few training samples are available, the quality of the proposed designs is acceptable. A more detailed analysis of the training phase shows that, for the proposed application, a data set with as little as 500 designs is sufficient to yield this accuracy. This is a considerable advantage of the proposed method over classical topology optimization: In our case, a single run of modal analyses yields sufficient information to quickly optimize designs for a wide range of requests. Conventional optimization schemes require typically hundreds of model evaluations to converge on a single design of for each new request.

Several prototypes with requested modal properties are manufactured. Through measurements, we are able to confirm that the real modal parameters are close to the predicted and simulated values. The error between the requested and achieved eigenfrequencies remains smaller than 6\%, and for the masses it is below 15\%. To our knowledge, no other work has tackled the multi-mode opimization of dynamic mass and frequency simultaneously. We finally show that this matching translates to accurate band gap generation in a beam-type LRM, both numerically and experimentally.

Different extensions or applications of the work may be explored in future research. First of all, tuned resonator design is not only relevant to metamaterials, but also to other resonance-based vibration control solutions \cite{Sun1995, Dedoncker2022a}. The generative approach outlined here could be easily modified to consider more input or output variables. Consideration of additional translational or rotational degrees of freedom in the host-resonator coupling naturally leads to an expansion of the (scalar) dynamic mass into a mass matrix, which relates the interface (generalized) motion and force vectors \cite{Girard208}. Rotational coupling in particular can be used to address a directional dependency of wave propagation, for example in orthotropic plates \cite {Giannini2023}. The cVAE allows to use very different output variables, not necessarily the structures' dynamic properties. It could potentially be applied to static mechanical features, thermal, acoustic, or optical properties.

On the machine learning level, more data-efficient training approaches, such as active learning, could be explored. It is so far not clear what the minimal amount of training samples should be, and how it is related to the amount of input and output parameters. User-friendliness and interactivity could be improved by providing more feedback and guidance. For instance, given the correlations in the training data, or the trained network's sensitivity to the inputs, it should be possible to estimate which combinations of performance attributes can be achieved and which design variables tend to have the largest impact on the performance.

\paragraph{Acknowledgments} This work was funded by a Swiss Data Science Center project grant (C20-07). We express our gratitude to Luis Salamanca, Aleksandra Anna Apolinarska, and Matthias Kohler for generously providing their code for the cVAE framework~\cite{Salamanca2023AugmentedStudy2}, which we trained the generative inverse design model with.

\bibliography{ref_vdb,references_cd}

\end{document}